\pdfoutput=1 
\documentclass{JINST}

\usepackage{amsmath,amssymb}
\usepackage[numbers,sort]{natbib}

\title{Improved background rejection in neutrinoless double beta decay experiments using a magnetic field in a high pressure xenon TPC}

\author{J. Renner$^a$\thanks{Corresponding author.},
A. Cervera$^a$, J.A. Hernando$^b$, A. Imzaylov$^a$, F. Monrabal$^a$, J. Mu\~noz$^a$, D. Nygren$^b$, and J. J. Gomez-Cadenas$^a$\\
\llap{$^a$}Instituto de F\'isica Corpuscular (IFIC), CSIC \& Universitat de Val\`encia,\\ 
Calle Catedr\'atico Jos\'e Beltr\'an, 2, 46980 Paterna, Valencia, Spain\\
\llap{$^b$}Instituto Gallego de F\'isica de Altas Energ\'ias (IGFAE), Univ.\ de Santiago de Compostela,\\ Campus sur, R\'ua Xos\'e Mar\'ia Su\'arez N\'u\~nez, S/N, 15782 Santiago de Compostela, Spain\\
\llap{$^c$}University of Texas at Arlington\\
 701 S. Nedderman Drive, Arlington, TX 76019, USA\\
E-mail: \email{jrenner@ific.uv.es}}

\newcommand{\bb}{\ensuremath{\beta\beta}}
\newcommand{\bbonu}{\ensuremath{0\nu\beta\beta}}

\newcommand{\mbb}{\ensuremath{m_{\bb}}}
\newcommand{\Tonu}{\ensuremath{T_{1/2}^{0\nu}}}

\newcommand{\Qbb}{\ensuremath{Q_{\bb}}}

\newcommand{\ckky}{\ensuremath{\mathrm{cts~keV^{-1}~kg^{-1}~yr^{-1}}}}

\newcommand{\XE}{\ensuremath{^{136}\mathrm{Xe}}}

\newcommand{\TL}{\ensuremath{^{208}\mathrm{Tl}}}
\newcommand{\BI}{\ensuremath{^{214}\mathrm{Bi}}}

\newcommand{\CHF}{\ensuremath{\mathrm{CH}_4}}
\newcommand{\CFF}{\ensuremath{\mathrm{CF}_4}}

\abstract{We demonstrate that the application of an external magnetic field could lead to an improved background rejection in neutrinoless double-beta (\bbonu) decay experiments using a high-pressure xenon (HPXe) TPC.  HPXe chambers are capable of imaging electron tracks, a feature that enhances the separation between signal events (the two electrons emitted in the \bbonu\ decay of \XE) and background events, arising chiefly from single electrons of kinetic energy compatible with the end-point of the \bbonu\ decay (\Qbb). Applying an external magnetic field of sufficiently high intensity (in the range of 0.5-1 Tesla for operating pressures in the range of 5-15 atmospheres) causes the electrons to produce helical tracks.  Assuming the tracks can be properly reconstructed, the sign (direction) of curvature can be determined at several points along these tracks, and such information can be used to separate signal ($0\nu\beta\beta$) events containing two electrons producing a track with two different directions of curvature from background (single-electron) events producing a track that should spiral in a single direction.   Due to electron multiple scattering, this strategy is not perfectly efficient on an event-by-event basis, but a statistical estimator can be constructed which can be used to reject background events by one order of magnitude at a moderate cost ($\sim$30\%) in signal efficiency. Combining this estimator with the excellent energy resolution and topological signature identification characteristic of the HPXe TPC, it is possible to reach a background rate of less than one count per ton-year of exposure.  Such a low background rate is an essential feature of the next generation of \bbonu\ experiments, aiming to fully explore the inverse hierarchy of neutrino masses.}


\begin{document}

\section{Introduction}\label{sec:intro}

Neutrinoless double beta decay (\bbonu) is a postulated very slow radioactive process in which two neutrons inside a nucleus transform into two protons, emitting two electrons. The discovery of this process would demonstrate that neutrinos are Majorana particles and that total lepton number is not conserved in nature, two findings with far-reaching implications in particle physics and cosmology \cite{GomezCadenas:2013ue, Cadenas_2012}.

After 75 years of experimental effort, no compelling evidence for the existence of \bbonu\ decay has been obtained. The current generation of experiments, with fiducial masses in the range of 100 kg of isotope and a total background count in the region of interest (ROI) around \Qbb\ of a few tens of counts per year, will barely explore the so-called degenerate hierarchy of neutrino masses ($m_1 \sim m_2 \sim m_3$). Exploring the full inverse hierarchy of neutrino masses would require a sensitivity to the neutrino Majorana mass of 20 meV, which in turn implies building detectors with fiducial masses in the range of one ton of isotope and a total background count in the ROI of, at most, a few counts per ton-year of operation. This is a tremendous experimental challenge which requires increasing the target mass of the current generation of experiments by typically one order of magnitude while at the same time decreasing the backgrounds by {\em at least} one order of magnitude (for a recent discussion and extensive bibliography on the subject, see \cite{Gomez-Cadenas:2015twa}).

One of the technologies currently being developed is that of high pressure xenon (HPXe) Time Projection Chambers (TPCs). In particular, the NEXT collaboration \cite{Gomez-Cadenas:2014dxa} is building a HPXe TPC capable of holding 100 kg of xenon enriched at 90\% in the \bb\ decaying isotope \XE. NEXT operates at 15 bar and uses electroluminescent (EL) amplification of the ionization signal to optimize energy resolution. The detection of EL light provides an energy measurement using 60 photomultipliers (PMTs) located behind the cathode (the \emph{energy plane}) as well as tracking  via a dense array of about 8,000 silicon photomultipliers (SiPMs) located behind the anode (the \emph{tracking plane}).

The NEXT experiment has completed the R\&D phase which was carried out with the large-scale prototypes NEXT-DEMO and NEXT-DBDM. The prototypes have measured an energy resolution which extrapolates to 0.5--0.7 \% FWHM at \Qbb. NEXT-DEMO has also shown the robustness of the topological signal \cite{Alvarez:2012xda,Alvarez:2012kua,Alvarez:2013gxa,Lorca:2014sra}. Currently, the NEXT collaboration is commissioning the first phase of the experiment, called NEW, at the Canfranc underground laboratory (LSC). NEW is a scaled-down version of NEXT-100, deploying 10 kg of isotope mass and 20\% of the sensors of NEXT-100 (12 PMTs in the energy plane and  1,800 SiPMs in the tracking plane).

The NEXT background model, which will be fully validated with NEW, predicts a background rate of $5 \times 10^{-4}$~\ckky\ in the ROI \cite{Nebot-Guinot:2014raa}. The energy resolution for NEXT-100 is assumed to be 0.7\% FWHM at \Qbb\ ($\sim$ 17 keV). NEXT-100 is scheduled to begin acquiring data in 2017. The experiment expects less than one count of background per 100 kg and year of exposure, and thus its sensitivity to \Tonu\ is not dominated by background subtraction and increases rapidly with exposure. The expected sensitivity to the \bbonu\ half-life is $\Tonu > 7 \times 10^{25}$~yr for a exposure of 300 kg$\cdot$yr. This translates into a \mbb\ sensitivity range of $[90-180]$~meV, depending on the nuclear matrix element. An important advantage of the HPXe technology is the fact that it can be extrapolated to ton-scale target masses.

A central feature of a HPXe TPC is the capability of imaging electron tracks, a feature that can be used to separate signal events (the two electrons emitted in a \bbonu\ decay) from background events (mainly due to single electrons with kinetic energy comparable to the end-point of the \bbonu\ decay, \Qbb). In this paper, we show that adding an external magnetic field of sufficiently high intensity (in the range of 0.5--1 Tesla for operating pressures in the range of 5-15 atmospheres) allows for improved separation of signal and background events provided that two key conditions are met. 
\begin{enumerate}
\item The spatial resolution needs to be sufficiently good (in the range of 2-3 mm). 
\item The electron track needs to be reconstructed with sufficient precision. 
\end{enumerate}

If the above two conditions are met, then a statistical estimator based on the average sign of the curvature of the electron track in the magnetic field can be constructed.  This estimator can be used to reject background events by one order of magnitude while maintaining a 60-80\% signal efficiency (depending on the pressure). Combining this estimator with the excellent energy resolution and topological signature characteristic of the HPXe-EL TPCs, we estimate that a future ton-scale detector will be able to reach less than one count background event per ton year of exposure, and thus meet the requirements needed to explore the inverse hierarchy.

This paper is organized as follows.  Section \ref{sec.topology} describes the reconstruction of the tracks of high-energy electrons produced in the dense gas of a high pressure TPC.  The simulation and analysis of such tracks in the presence of an external magnetic field is described in section \ref{sec.magmotion}.  In section \ref{sec.curvature}, the analysis procedure is discussed, and the Monte Carlo data sample to which it is applied is described in section \ref{sec.track}.  Finally, the main results are shown in sections \ref{sec.results} and \ref{sec.improvedanalysis}, and their implications are discussed in section \ref{sec.outlook}.

\section{Imaging tracks in a HPXe TPC}
\label{sec.topology}

\begin{figure}[!htb]
\centering
\includegraphics[width= 0.95\textwidth]{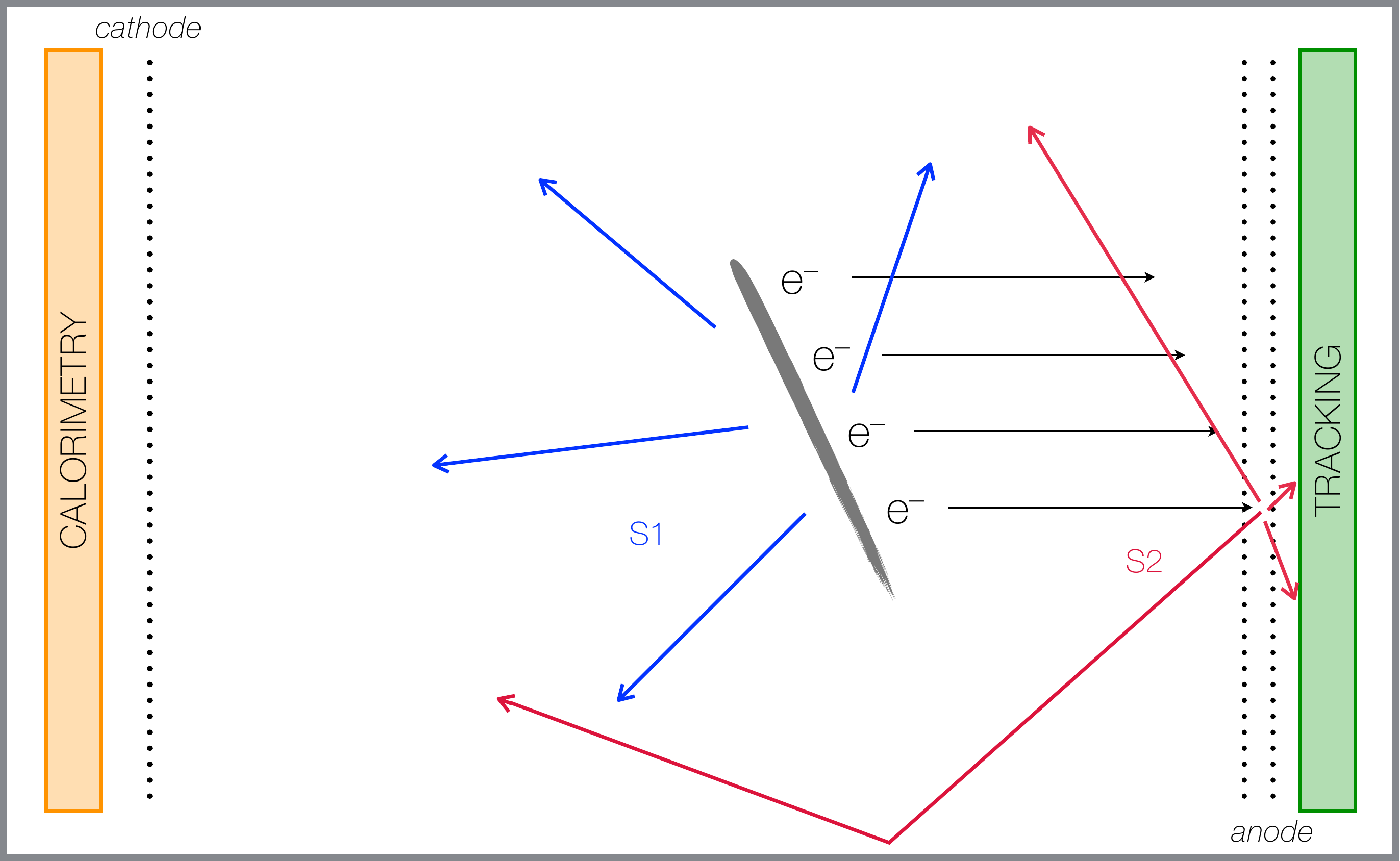}
\caption{Principle of operation of an asymmetric HPXe TPC with EL readout.} \label{fig.SS}
\end{figure}

Figure \ref{fig.SS} shows the principle of operation of an asymmetric HPXe TPC using proportional electroluminescent (EL) amplification of the ionization signal (as is the case for NEXT-100). The detection process involves the use of the prompt scintillation light ($S_1$) from the gas as the start-of-event time, and the drift of the ionization charge to the anode by means of an electric field ($\sim0.3$ kV/cm at 15 bar) where secondary EL scintillation ($S_2$) is produced in a narrow region defined by two highly transparent meshes (called the EL gap or ELG).  High voltages are applied to the two meshes to establish an electric field of $\sim20$ kV/cm at 15 bar in this region. The detection of EL light provides an energy measurement using PMTs in the case of NEXT-100 located behind the cathode (the \emph{energy plane}). The reconstruction of the track topology is carried out with a dense dense array of SiPMs located behind the anode (the \emph{tracking plane}). The $x$-$y$~coordinates are found using the information provided by the tracking plane, while $z$~is determined by the drift time between the detection of $S_1$~and $S_2$. For each reconstructed space point, the detector also measures the energy deposited. Thus, the track is imaged as a collection of hits. Each hit is defined by a 3D space coordinate and by an associated energy deposition, as (x,y,z,E).

\begin{figure}[!htb]
\centering
\includegraphics[width= 0.95\textwidth]{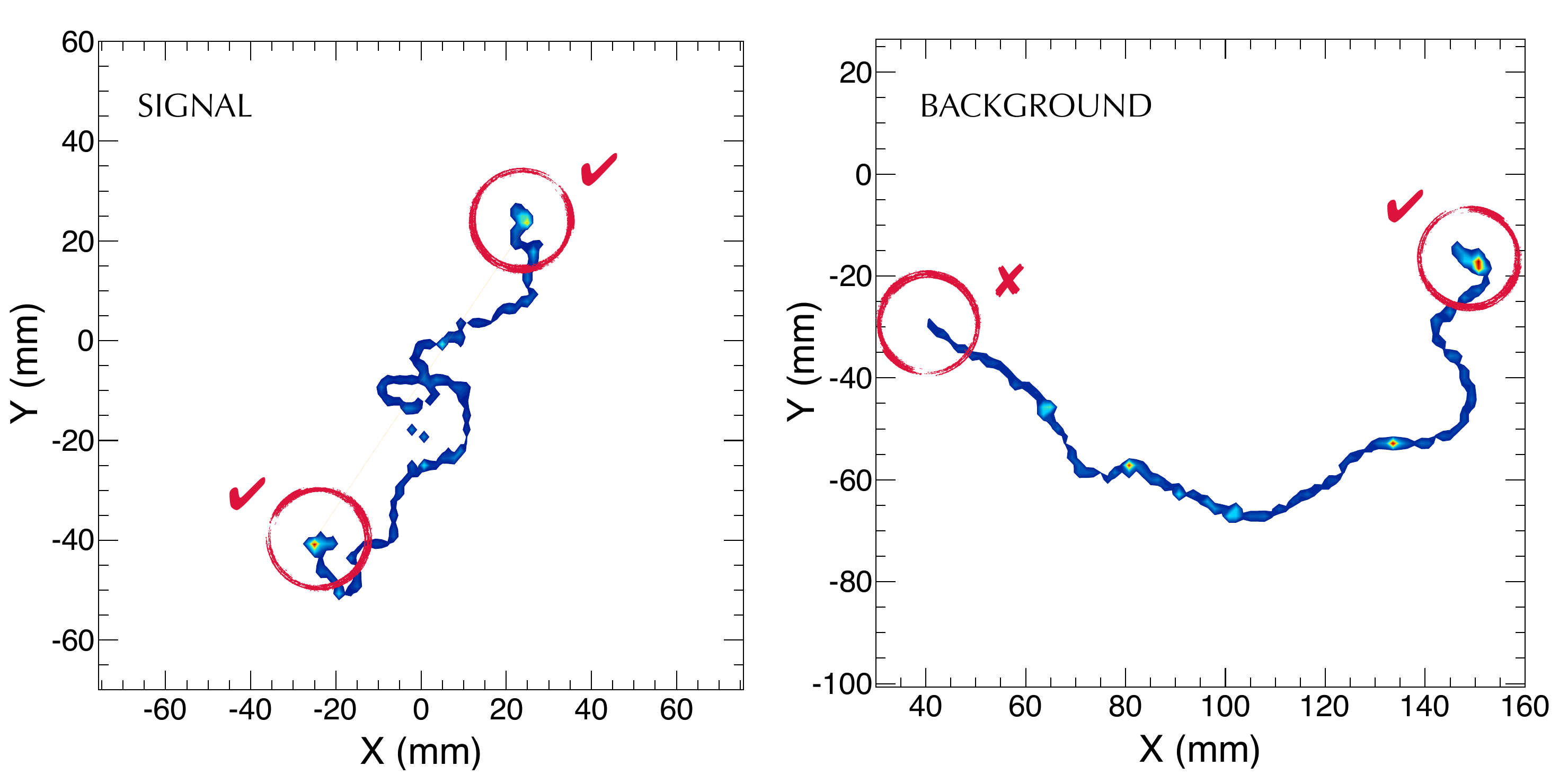}
\caption{Monte Carlo simulation of a signal (\bbonu) event (left) and a  background event (right) in xenon gas at 15~bar. The color corresponds to energy deposition in the gas. The signal consist of two electrons emitted from a common vertex, and thus it features large energy depositions  (blobs) at both ends of the track. Background events are, typically, single-electron tracks (produced by photoelectric or Compton interactions of high energy gammas emitted by \BI\ or \TL\ isotopes), and thus feature only one blob.} \label{fig.ETRK2}
\end{figure}

\begin{figure}[!htb]
\centering
\includegraphics[width=0.45\textwidth]{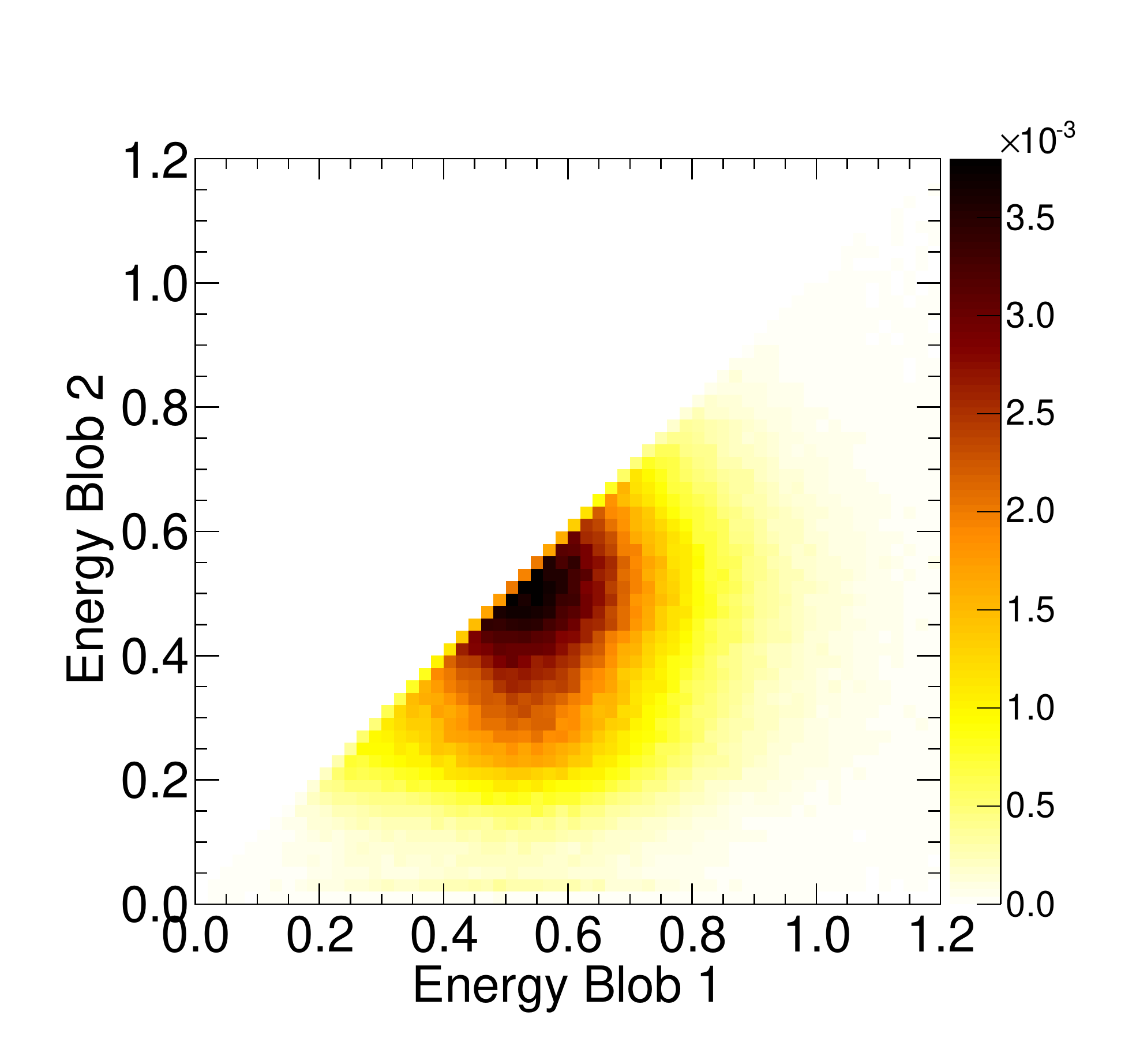}
\includegraphics[width=0.45\textwidth]{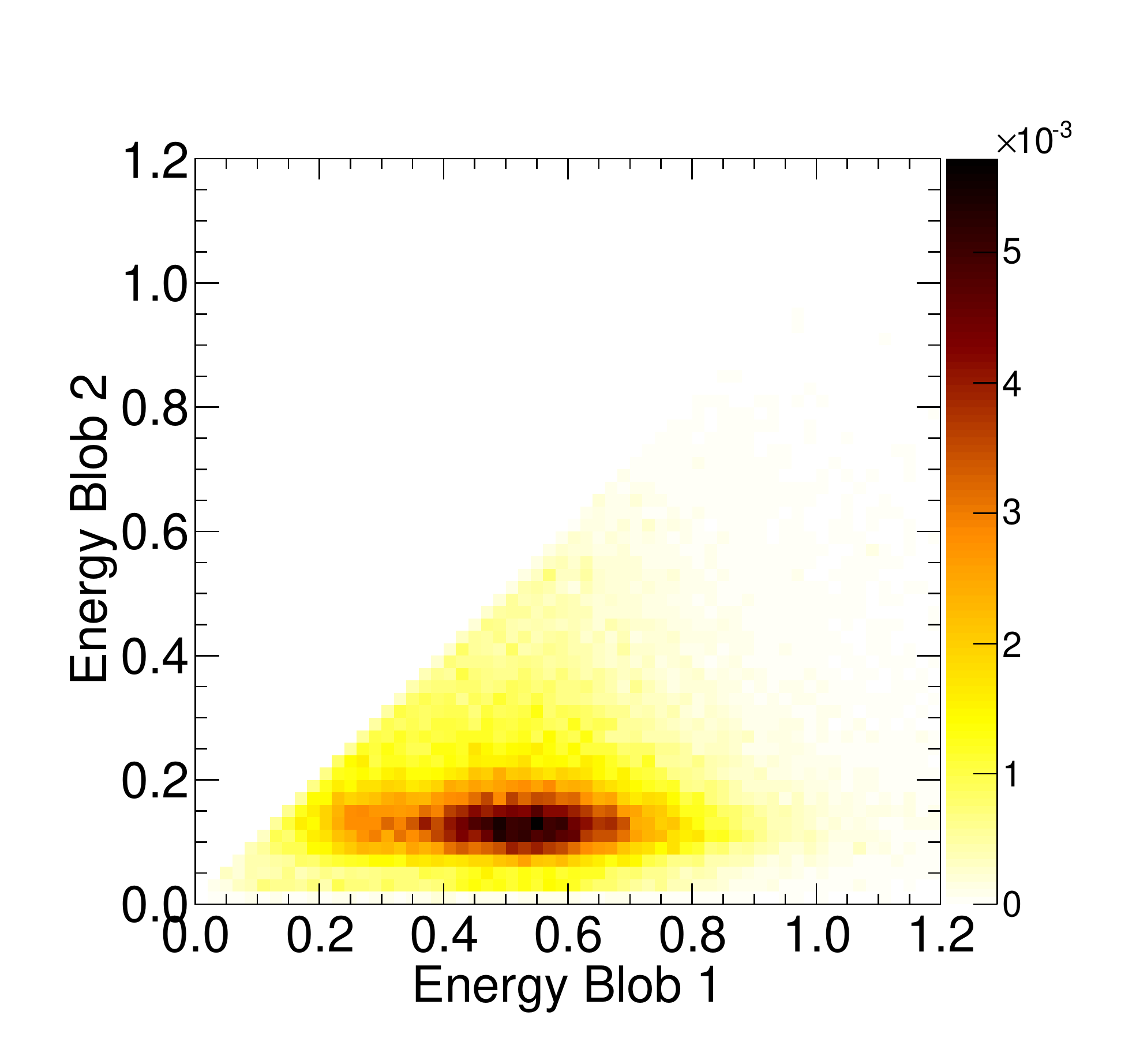}
\caption{Probability distribution of signal (left) and background (right) events in terms of the energies of the end-of-track blobs. The blob labelled as `1' corresponds to the more energetic one, whereas `blob 2' corresponds to the less energetic of the two. In a signal event, the blobs have, on average, the same energy. In a background event, blob 1 has an energy similar to that of a signal event while the energy of blob 2 is very small.} \label{fig.BLOBS}
\end{figure}

Double beta decay events leave a distinctive topological signature in a HPXe TPC: a continuous track with larger energy depositions (\emph{blobs}) at both ends due to the Bragg-like peaks in the d$E$/d$x$ of the stopping electrons (figure \ref{fig.ETRK2}, left). In contrast, background electrons are produced by Compton or photoelectric interactions, and are characterized by a single blob in one of the ends (figure \ref{fig.ETRK2}, right). Reconstruction of this topology using the tracking plane provides a powerful means of background rejection. For each track, the energy in both extremes of the track is measured and labelled as $E_{b1}$~ (the energy of the most energetic energy deposition), and $E_{b2}$~ (the least energetic energy deposition). In a signal event, $E_{b1} \sim E_{b2} $, while for background events $E_{b1} >> E_{b2} $. Figure \ref{fig.BLOBS} shows how this feature can be used to separate signal from background. 

In this study we compare the single and double-electron track signatures in the presence of an external magnetic field and quantify the additional ability to reject single-electron events that is gained by examining the curvature of the tracks induced by the field.  We note that the ability to conduct a detailed study of the reconstructed track is highly dependent on the spatial resolution with which the track can be reconstructed in the detector {\em and} on developing a reconstruction algorithm with minimum bias to find the track extremes and track orientation. 

Spacial resolution, in turn, depends on:
\begin{enumerate}
\item {\em Tracking plane design:} this includes the pitch of the SiPMs in the tracking plane as well as the SiPM response. Indeed, the use of SiPMs with very low dark current and high gain allows one to determine the location of an event by weighting the position of each sensor with the light recorded, thus improving dramatically the ``digital'' resolution, which goes as $\sim$pitch/$\sqrt{12}$. The digital point resolution corresponding to a pitch of 10 mm (NEXT-100) is 10$/\sqrt{12} \sim 3$~mm. Using weighted information (e.g, local barycenter algorithms), the point resolution improves to about 1 mm. 
\item {\em The width of the EL region:}
The non-zero width of the EL gap adds an extra resolution term which goes like $w/\sqrt{12}$~where $w$~is the width of the grid. For the NEXT-DEMO prototype, $w\sim 5$, so this effect would add a resolution term of 1.5 mm. For NEXT-100, $w\sim 3.5$~reducing the resolution to 1 mm. 

\item {\em Diffusion of the drifting electron cloud:}  both longitudinal and transverse diffusion are high in pure xenon (of the order of $10$~mm/$\sqrt{\rm{m}}$). On the other hand, work in progress within the NEXT collaboration\footnote{``A homeopathic cure to large diffusion in pure xenon'', the NEXT collaboration, paper in preparation} suggests that adding small amounts of cooling gases such as \CHF\ or \CFF\ to pure xenon (at the level of 0.1 \% of \CHF\ or 0.01\% of \CFF\ reduce both transverse and longitudinal diffusion to some $2.0$~mm/$\sqrt{\rm{m}}$.
\end{enumerate}

Concerning reconstruction algorithms, the collection of energy depositions (``hits'') that are provided by the tracking plane must be organized into an ordered track in which the beginning and end of the track are correctly identified.  This is a difficult task due to the wide variety of track geometries that can result from multiple scattering, and it must be noted that the results of this study assume it is done correctly for all events.  The NEXT collaboration is currently working on a reconstruction algorithm capable of properly ordering the hits with minimum bias towards specific geometries. 

To summarize, it appears possible to achieve a resolution of $\sim$2-3 mm in x-y-z, and to reconstruct tracks with minimum bias as discussed before. In the remainder of this paper we assume both conditions. Our Monte Carlo study smears the individual hits by 2-3 mm and assumes that the reconstruction algorithm provides the correct ordering of the track. 

\section{Electron tracks propagating in high pressure gas inside a magnetic field}\label{sec.magmotion}

A particle of charge $q$ moving at a velocity $\mathbf{v}$ in the presence of a magnetic field $\mathbf{B}$ is acted upon by a force
%
$\mathbf{F} = q(\mathbf{v} \times \mathbf{B})$.
%
This force will cause an electron (propagating in vacuum) to execute a helical motion in the magnetic field such that if the thumb of the right hand is positioned along the direction of the field line $\mathbf{B}$ the electron will rotate in the direction in which the fingers curl when closed around the field line.  The frequency of rotation about the field line is known as the cyclotron frequency $\omega_{\mathrm{cyc}} = qB/m$, where $B$ is the magnitude of the magnetic field $B = |\mathbf{B}|$ and $m$ is the mass of the charged particle. The radius of curvature $r$ of the trajectory is directly proportional to the electron transverse momentum $p_{T}$ and to the inverse of the magnitude of the magnetic field, $r = p_{T}/qB$.

If the direction of the applied magnetic field is known, the curvature of the track can be calculated to determine whether the component of the electron velocity along the magnetic field is parallel or antiparallel to the field.  Assuming that the magnetic field is directed along the z-axis, we are interested in the curvature of the track in the $x$-$y$ plane as it progresses in $z$.  The curvature $\kappa$ in the $x$-$y$ plane can be calculated for a track parameterized by the coordinate $z$ as

\begin{equation}\label{eqn_curv}
\kappa = \frac{(dx/dz)\cdot(d^2y/dz^2) - (dy/dz)\cdot(d^2x/dz^2)}{\Bigl[(dx/dz)^2 + (dy/dz)^2\Bigr]^{3/2}}.
\end{equation}

With this definition, an electron traveling in the direction of the magnetic field will rotate around the field lines with positive curvature, while an electron traveling opposite the direction of the magnetic field will spiral with negative curvature.  

The curvature, however, will be of the opposite sign if the track orientation is not properly identified in the calculation (i.e., if $dz$ is of the wrong sign).  Thus, when calculating the curvature of a single-electron track, one would expect $\kappa > 0$ for $dz > 0$ and $\kappa < 0$ for $dz < 0$ given that $dz$ is always in the direction of the electron velocity.  However, for a \bbonu\ ``double electron'' track (defined as the track between the two blobs that mark the start and end of the trajectory), taking one of the extremes to be the beginning of the track and the other to be the end will lead to a calculation of $\kappa$ assuming the wrong track orientation for one of the two electrons, as the vertex at which the reaction occurred is found somewhere on the interior of the track.  Therefore one expects to find $\kappa > 0$ for $dz < 0$ and $\kappa < 0$ for $dz > 0$ for a significant fraction of the track.  This difference in the behavior of the calculated curvature of reconstructed tracks can be used to distinguish single-electrons from \bbonu\ double electrons.

One serious limitation to this strategy is the presence of electron multiple scattering (MS). This is a process by 
which an electron is scattered repeatedly while traveling through the dense gas, resulting in deviations of its
path by significant angles. The process can be described approximately by considering the angle of deflection projected on the $x$ and $y$ planes when scattered through a thickness of xenon $x$ (assuming that the electron is traveling along an axis in the $\hat{\mathbf{z}}$ direction). Each of these angles $\theta_x$ and $\theta_y$ can be modeled by a gaussian distribution (Moli\`{e}re theory) as

\begin{equation}\label{eqn_mscat}
\sigma^{2}(\theta_{x,y}) = \frac{13.6\,\,\mathrm{MeV}}{\beta p}\sqrt{dz/L_{0}}\bigl[1 + 0.038\ln(dz/L_{0})\bigr],
\end{equation}

\noindent for an electron with momentum $p$ in MeV/c before scattering and beta factor $\beta$ (where $c = 1$).  $L_{0}$ is the radiation length, a property of the medium. 

In the absence of MS the transverse momentum of an electron could be directly determined by the track
curvature. Consequently, a single-electron track---for which the momentum is greatest at one end of the
track and decreases toward the other end---would be clearly distinguishable
from a double-electron track consisting of two electrons traveling in opposite directions originating at some vertex in the middle of the track.  Unfortunately, in a HPXe TPC multiple scattering is high, resulting in large errors that complicate the direct measurement of the magnitude of the curvature. The sign of the curvature, however, is much less affected by MS.
Consequently we choose the sign of the curvature, rather than its magnitude, as our discriminating variable.  

\section{Determination of the Track Curvature}\label{sec.curvature}
Given a track of electron ionization defined as a collection of hits containing position and energy information, $(x,y,z,E)$, the curvature at each hit in the track is calculated by numerical computation of the derivatives $dx/dz$, $dy/dz$, $d^2x/dz^2$, and $d^2y/dz^2$.  From these the curvature $\kappa$ is calculated at each point.  Since we do not have $x$ and $y$ as a function of $z$ but rather $x$, $y$, and $z$ as a function of hit number $n$, we can calculate the derivatives $x' \equiv dx/dn$, $y' \equiv dy/dn$, $z' \equiv dz/dn$ using the chain rule as $\frac{dx}{dz} = x'/z'$, and

\begin{equation}
\frac{d^2x}{dz^2} = \frac{x'' - z''(dx/dz)}{(z')^2}.
\end{equation}

The expressions for $dy/dz$ and $d^2y/dz^2$ can be obtained by replacing in the above $x \rightarrow y$.  Note that outliers may need to be removed from the resulting arrays of first and second derivatives due to points between which the z-coordinate changes very little.  To ensure more stable values of the derivatives, an outlier removal procedure is applied to all derivatives and second derivatives computed which consists of iteratively calculating the mean and variance $\sigma$ of each array, replacing any value that lies outside of $5\sigma$ of the mean value with the average of the two nearest values in the array, and continuing this procedure until the calculated variance is no longer less than the value of the variance calculated in the previous iteration.

The curvature calculated using each pair of points is then corrected as follows: if for the two points $z_2 < z_1$, that is 
$dz < 0$, the curvature is multiplied by -1 (see section \ref{sec.magmotion}).  Note that the outlier removal procedure described above is also applied to the calculated curvature array.  

The discriminant variable is obtained by condensing the curvature information for each track into a single numerical value indicative of the nature of the track.  First a curvature sign array is created 
consisting of values of either $+1$ or $-1$ depending on the sign of each value in the calculated curvature array.  The 
numerical value, which we will
call the {\em curvature asymmetry factor (CAF)}, is defined as the average of the curvature sign array using elements in the first half of 
the track minus the average of the curvature sign array using elements in the second half of the track:

\begin{equation}\label{eqn_assym}
\phi_{C} = \frac{1}{N/2}\Biggl(\sum_{k=0}^{N/2-1}\mathrm{sgn}(\kappa_{k}) - \sum_{k=N/2}^{N}\mathrm{sgn}(\kappa_{k})\Biggr).
\end{equation}

\noindent Note that if the number of hits in the track is odd, the first term includes the first $(N-1)/2$ hits and
the second term includes the remaining $(N-1)/2 + 1$ hits.  Though due to multiple scattering and imperfect
reconstruction resolution, this factor is not a perfect discriminator, statistically it can be used to separate 
single-electron and double-beta events.

To further illustrate this idea, we consider a simple toy Monte Carlo in which tracks are generated in steps of 10 keV energy
deposition using the NIST \cite{NIST_mac} $dE/dx$ curve and equation \ref{eqn_mscat} to model multiple scattering.  For each energy 
deposition $\Delta E = 10$ keV, the electron is propagated along a path, calculated according to the force exerted
by the magnetic field, of length $\Delta x = \int_{E}^{E-\Delta E}(dE/dx)^{-1}dE$.   The
initial direction of travel is assigned at random and is updated after each step $\Delta x$ 
by calculating a random $\theta_x$ and $\theta_y$ using equation \ref{eqn_mscat} and
assuming the direction of travel is along the z-axis.  A single-electron track is generated in this way 
assuming an initial energy of 2.447 MeV, to simulate background events.  For ``signal'' events, two 
single-electron tracks are generated from a common vertex, each with an initial energy of $2.447/2 = 1.2235$ MeV. Notice that this is not realistic, since the two electrons produced in \bbonu\ events share their energy only statistically. Event-by-event, each electron can have an energy that varies between 0 and \Qbb\, with the other electron taking the rest. This makes it much harder, in practice, to separate \bbonu\ events from single electrons than in the naive case assumed here for illustration purposes.

\begin{figure}[!htb]
	\includegraphics[scale=0.48]{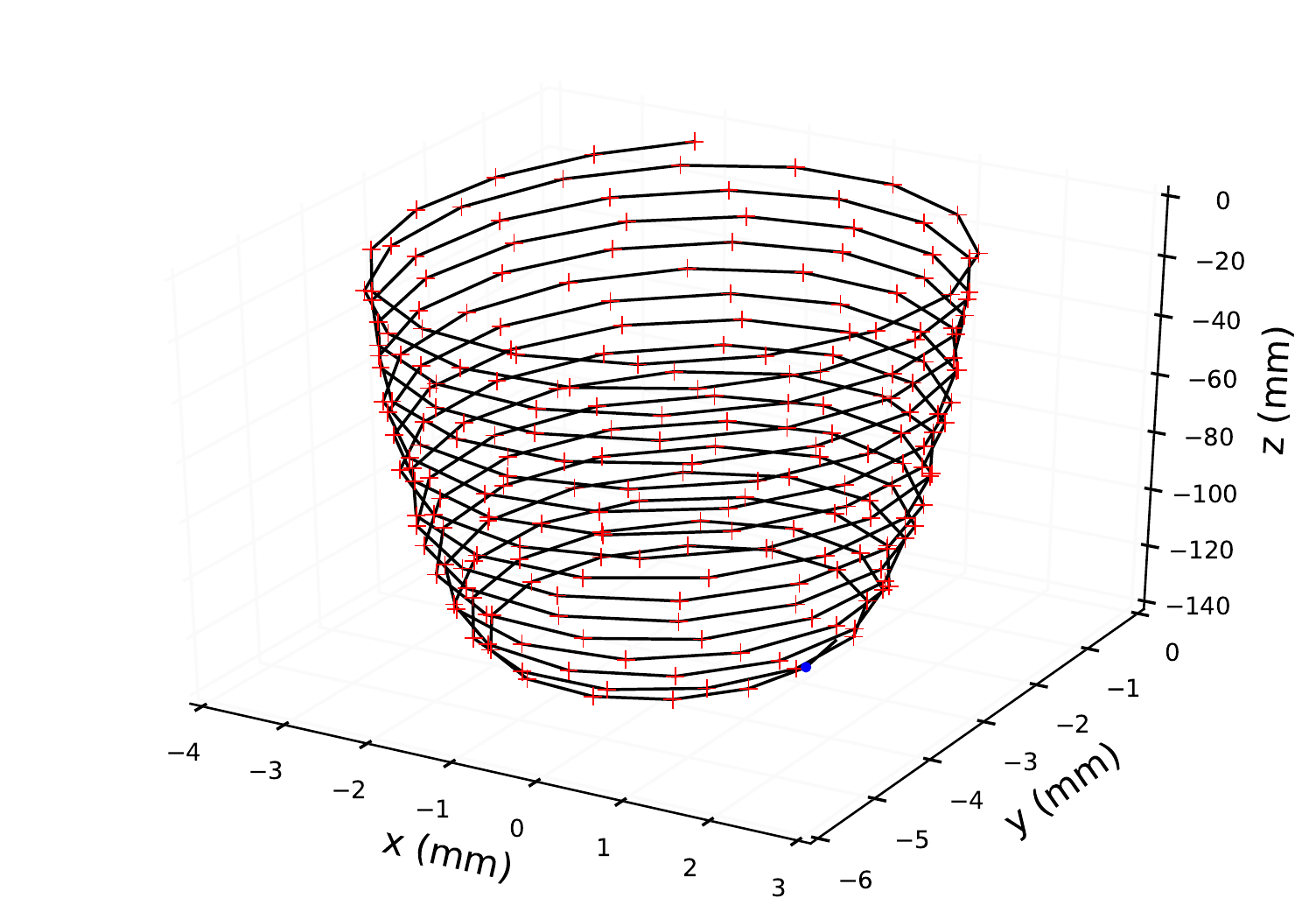}
	\includegraphics[scale=0.48]{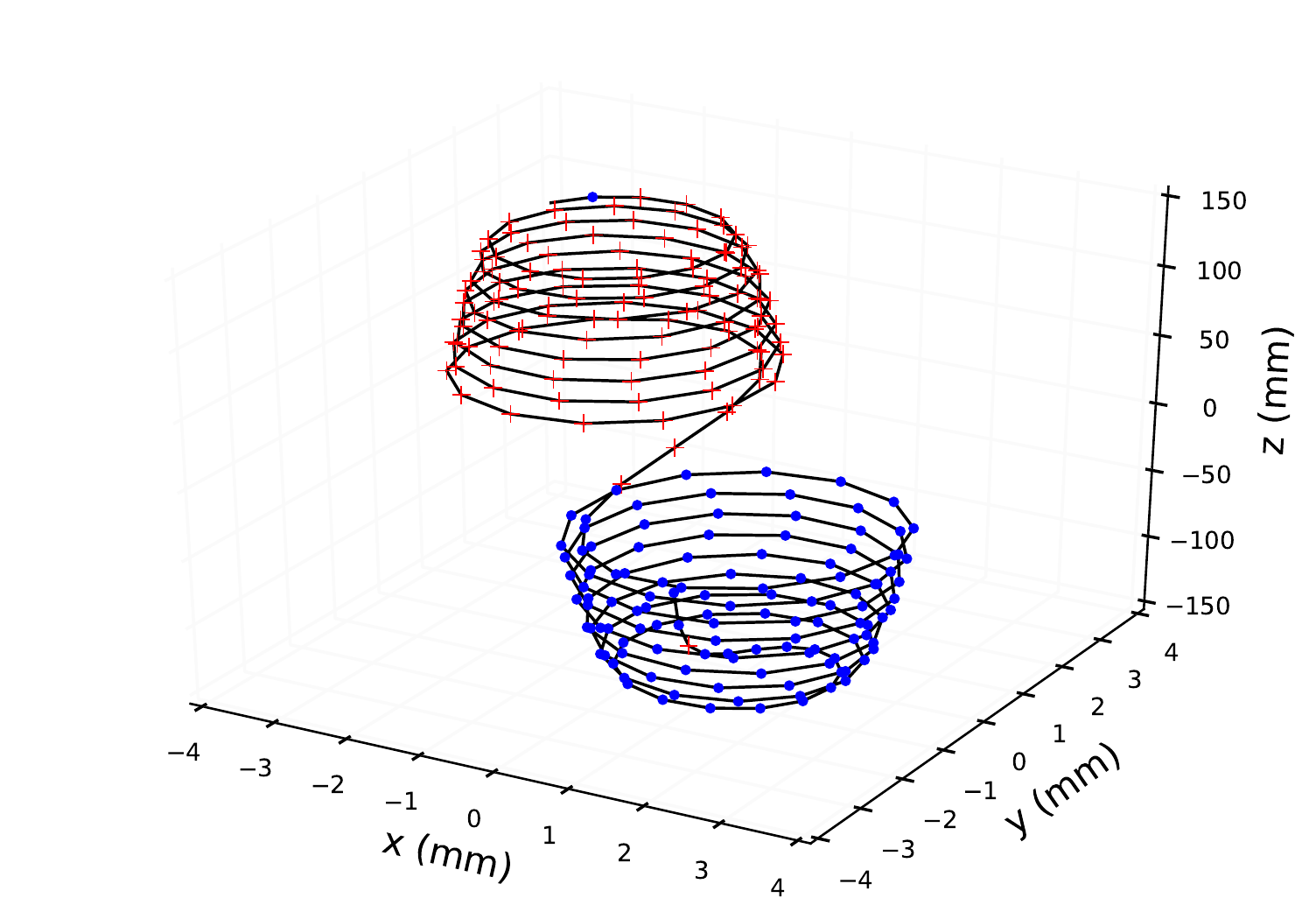}
	\includegraphics[scale=0.48]{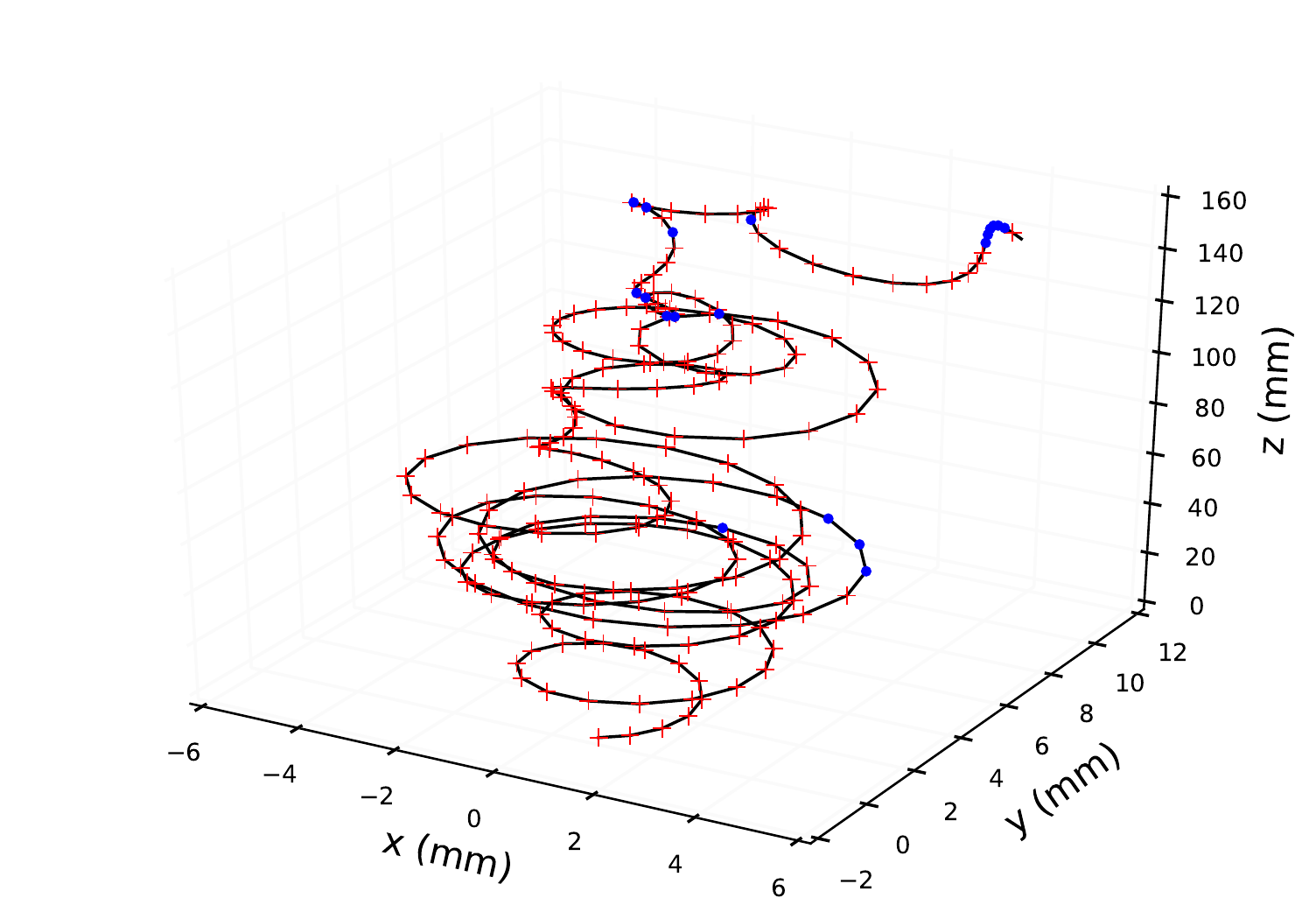}
	\includegraphics[scale=0.48]{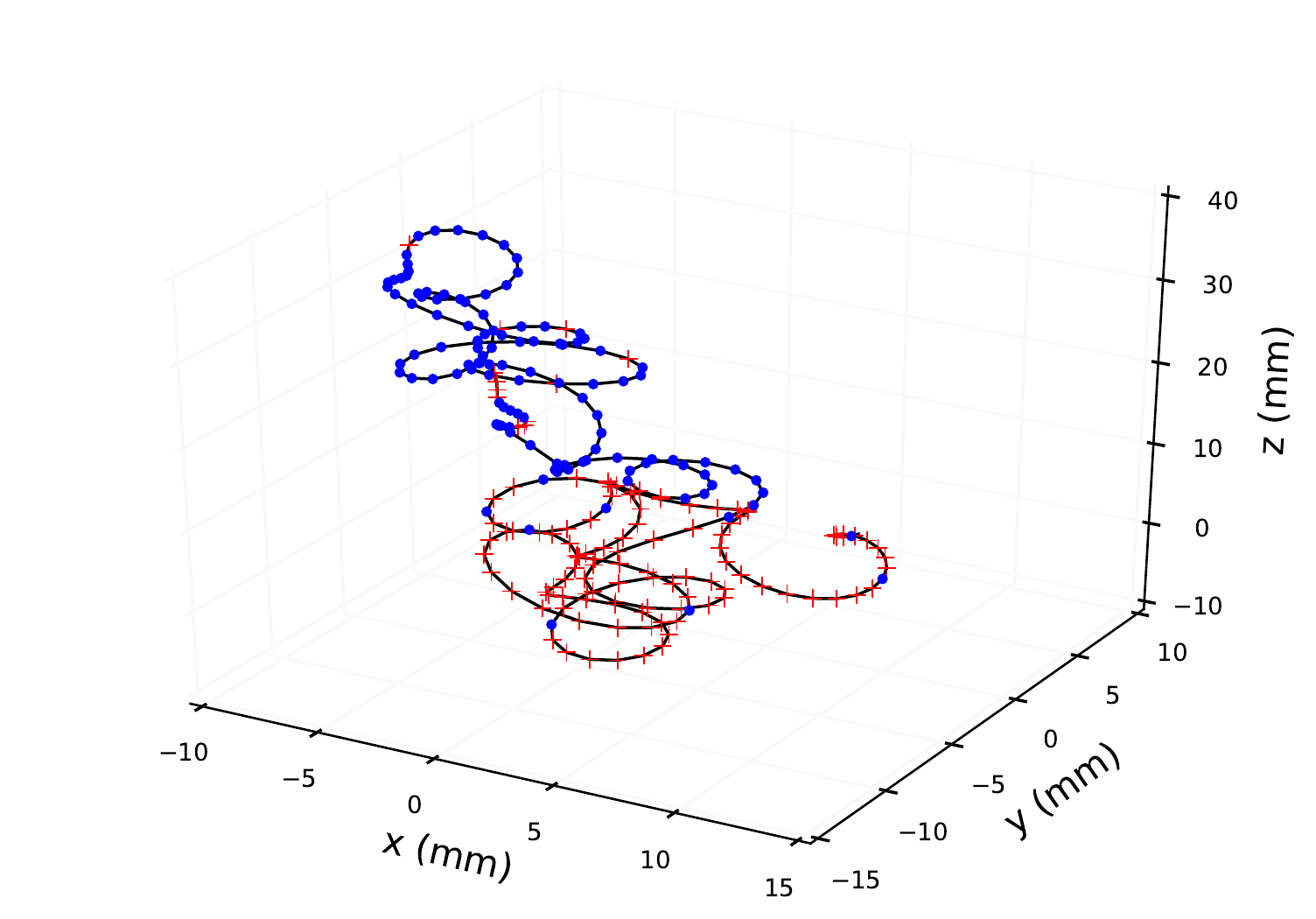}
	\includegraphics[scale=0.48]{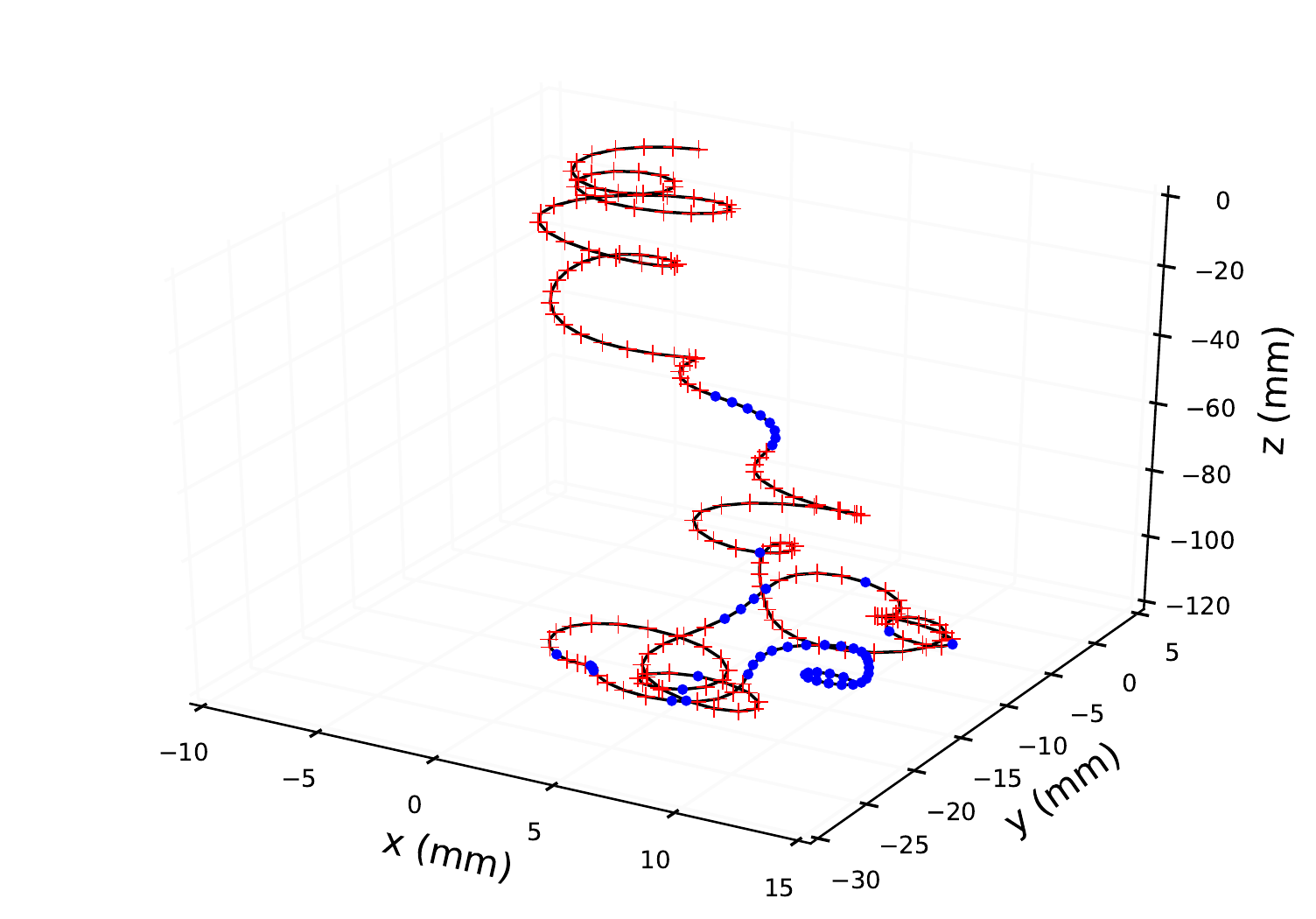}
	\includegraphics[scale=0.48]{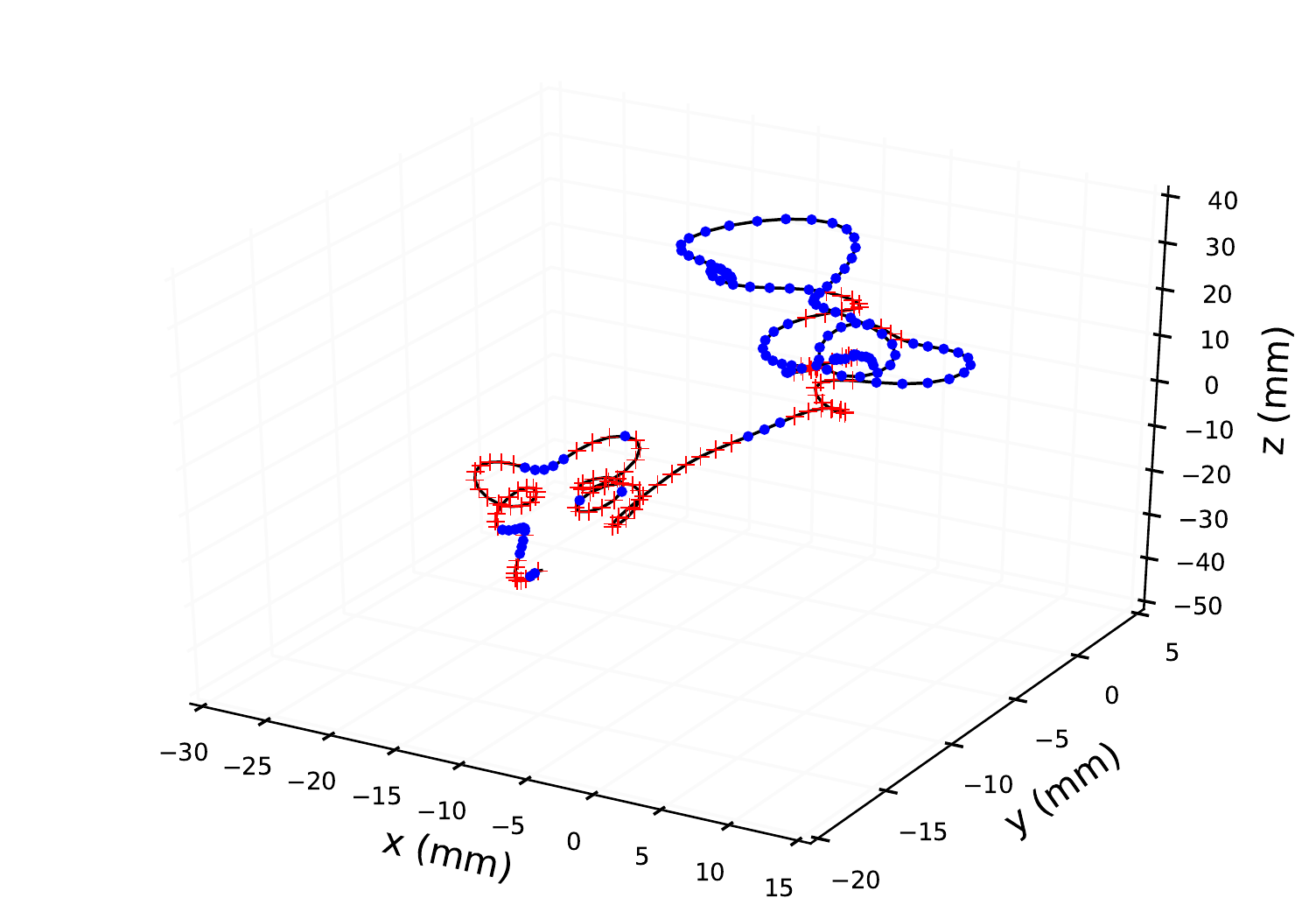}
	\caption{\label{fig_trkcurv_cf}Calculated curvature sign at each point along the track for a single-electron event (left) and a (simplified, see text) double-beta event (right).  The tracks were generated with no multiple scattering (top), the 
	standard amount of multiple scattering (middle) and twice the standard amount of multiple scattering 
	(bottom).  The red $+$ markers indicate positive curvature, while the blue dots indicate negative curvature.}
\end{figure}

Our toy Monte Carlo allows us to study the influence of (the gaussian component of) multiple scattering present in the simulated events in a
straightforward manner by changing the pre-factor in equation \ref{eqn_mscat} from 13.6 MeV to a lower or higher value.  
Figure \ref{fig_trkcurv_cf} shows several signal and background tracks and their calculated curvature sign 
simulated with different multiple scattering pre-factors corresponding to no multiple scattering (0 MeV), the 
standard amount of multiple scattering (13.6 MeV) and twice the standard amount of multiple scattering (27.2 
MeV).  

\begin{figure}[!htb]
	\includegraphics[scale=0.45]{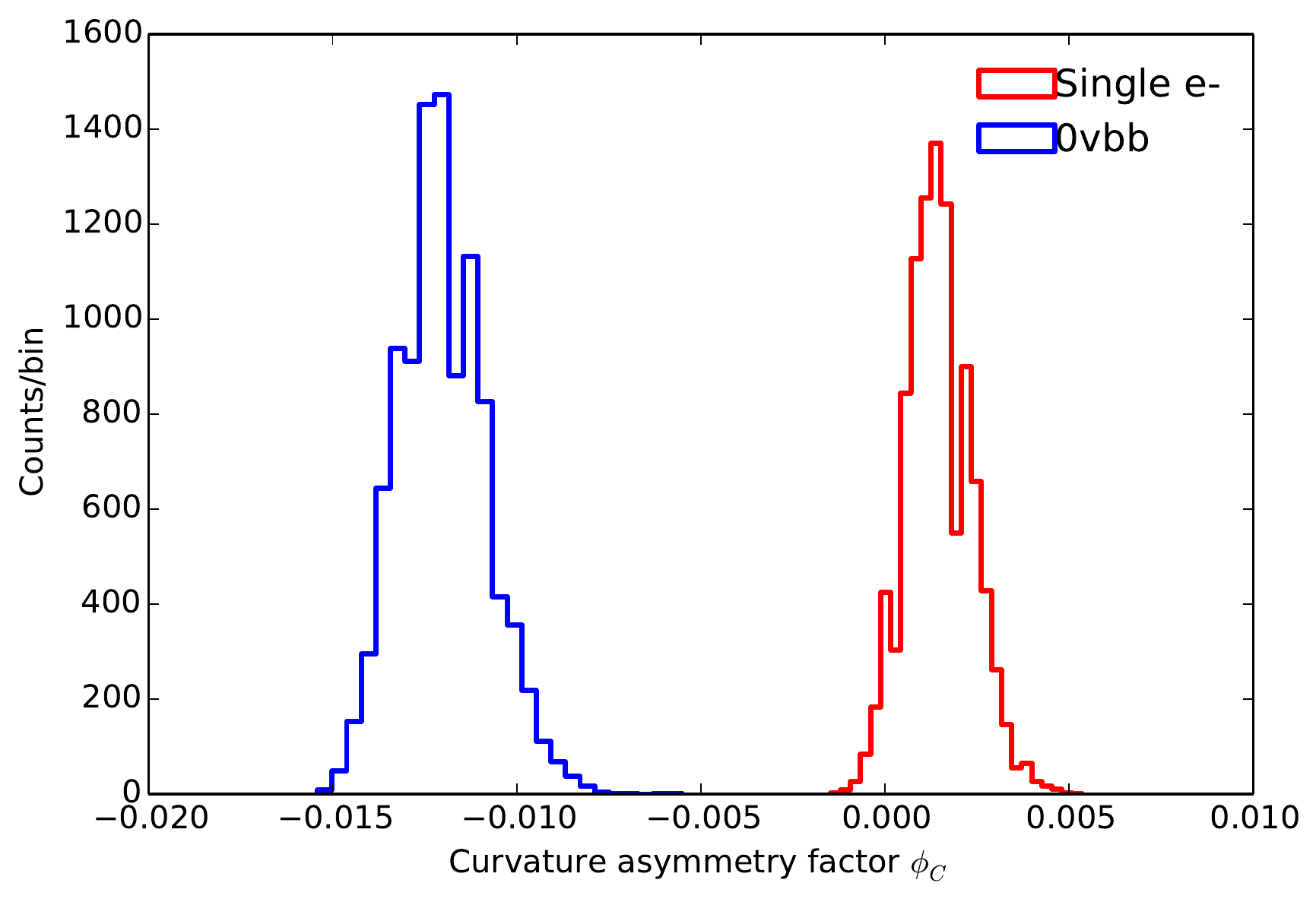}
	\includegraphics[scale=0.45]{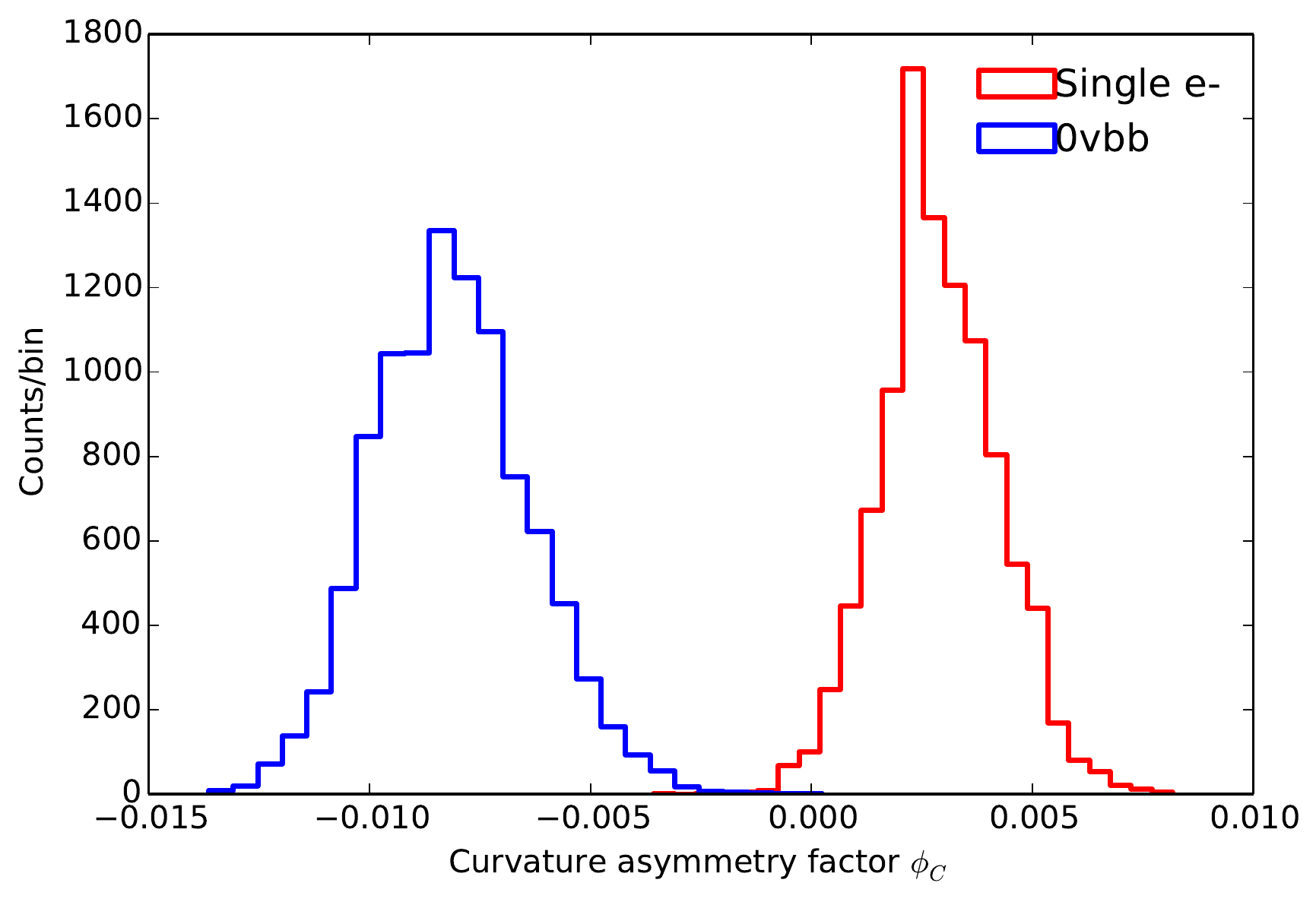}
	\caption{\label{fig_assym_cf} CAF calculated for $10^4$ single-electron and double-electron tracks generated with the standard amount of multiple scattering (left) and twice the standard amount of multiple scattering (right).  Note that though the background rejection capability appears almost perfect based on the results of this simple Monte Carlo, other factors such as the reconstruction resolution and non-equal energy sharing between the two electrons in \bbonu\ decay degrade the discrimination power and are not taken into account here.}
\end{figure}

Increasing multiple scattering  makes the background rejection procedure more difficult, as the random motion
of MS often works against the curvature introduced by the magnetic field and quantified in
equation \ref{eqn_assym}. However, as shown in figure \ref{fig_assym_cf} the CAF separates single-electron and double-electron tracks perfectly, even when the amount of MS is doubled. Naturally, this result is too optimistic, as already mentioned, for several reasons: the toy MC only includes the gaussian component of MS, neglecting important non-gaussian effects (included in the Geant-4 simulation), the point resolution is assumed perfect, and the "double-electrons" share exactly half of the energy. However, the good separation shown in figure \ref{fig_assym_cf} demonstrates that the proposed procedure is sound. 

\section{Data sample}\label{sec.track}
The primary simulation-based study consisted of analysis of Monte Carlo datasets generated in a large virtual box of high pressure xenon gas using GEANT4 \cite{GEANT4}. Such an arrangement emulates events contained in the fiducial volume of NEXT-100. For various configurations of gas pressure and magnetic field, $10^4$ signal (\bbonu) events and $10^4$ background events (single electrons of kinetic energy equal to \Qbb) were generated.  

Each simulated track was recorded as a series of hits consisting of a location $(x,y,z)$ and a deposited energy $E$.  Space coordinates were grouped in a single continuous track constructed by adding together all the hits left by the electron (in the case of single electron events) or the two electrons (in the case of \bbonu\ events). This procedure emulates the pattern recognition in a HPXe TPC, which reconstructs as a single track both single and double electrons (see figure \ref{fig.ETRK2}), and assumes that the hits are reconstructed in the proper order. Furthermore, the hits were smeared to simulate point resolution. Before proceeding with further analysis, the summed energy of all hits recorded in the track was required to be at least 2.4 MeV. This cut corresponds to demanding the event energy to be within the ROI. 

\begin{figure}[!htb]
	\centering
	\includegraphics[scale=0.48]{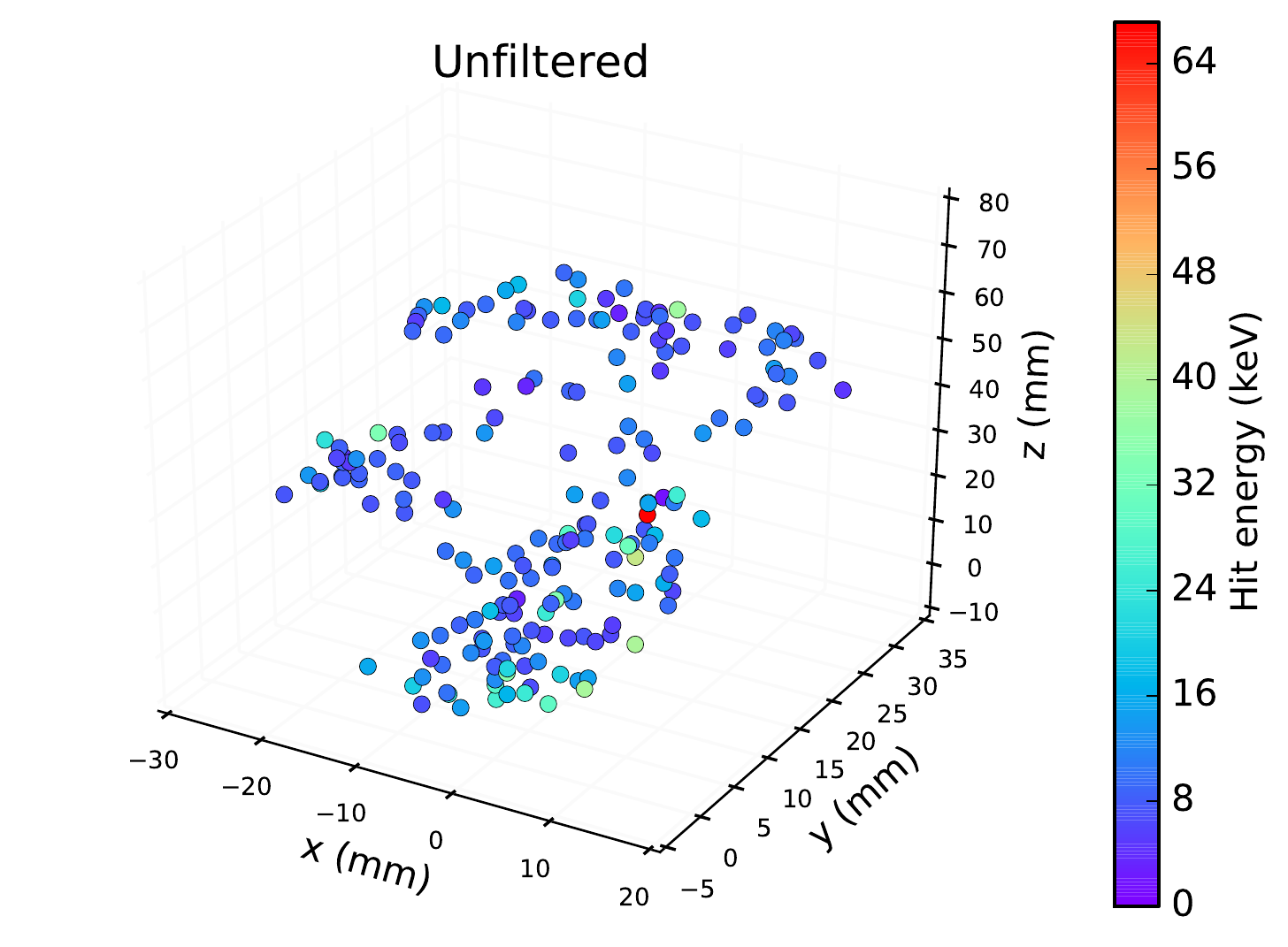}
	\includegraphics[scale=0.48]{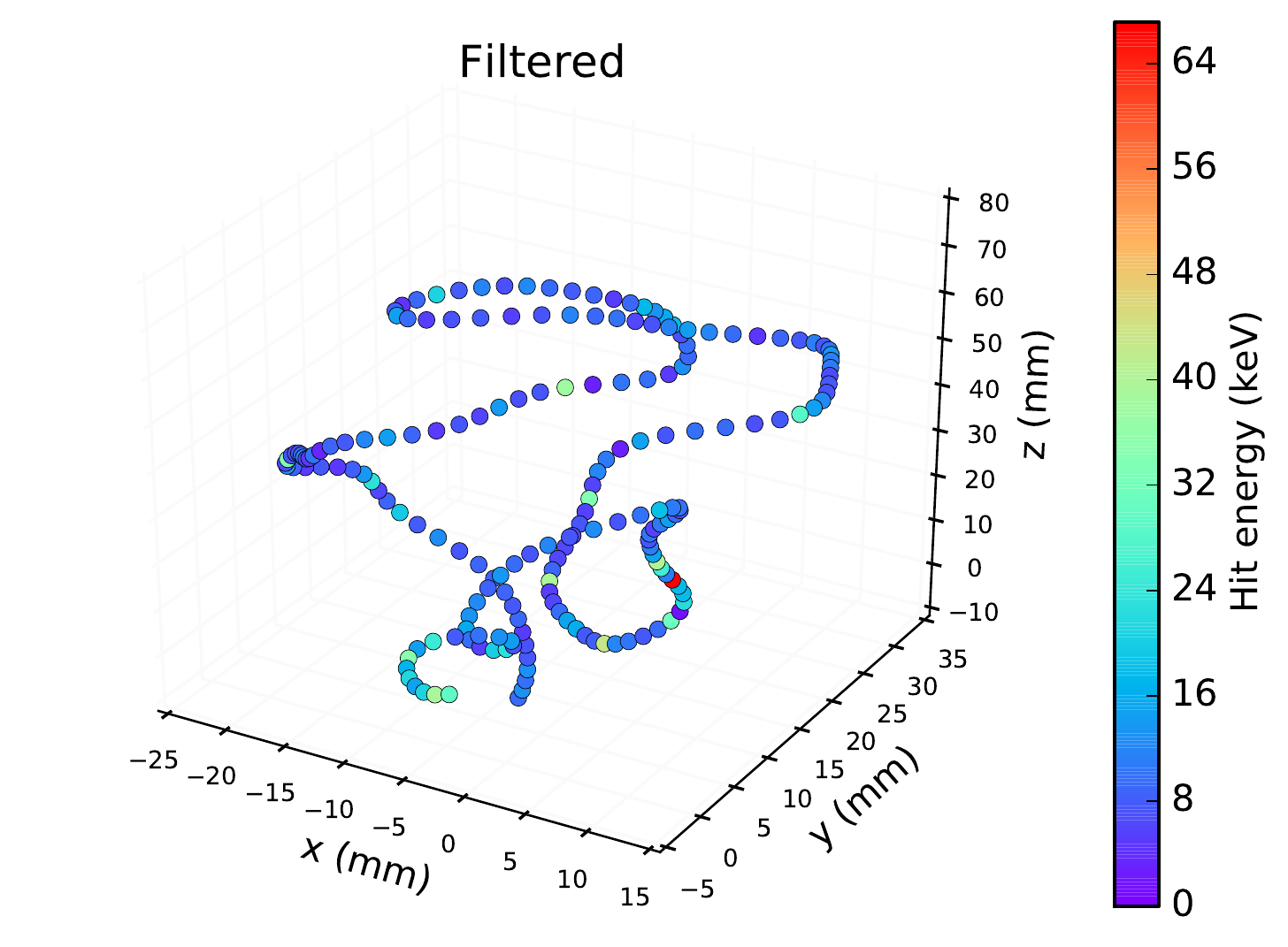}
	\caption{\label{fig_flt}An example of a single-electron track before (left) and after (right) the application of a lowpass filter.  The filter applied to this track is shown in appendix \protect\ref{app:FIR}, figure \protect\ref{fig_FIR}.}
\end{figure}

The x, y, and z-coordinates of the track were then separated into individual arrays, and a lowpass 
filter was applied to each array.  This served to smooth the track and facilitates the calculation of 
the derivatives.  Figure \ref{fig_flt} shows an example of a track before and after the filtering procedure, the
details of which are discussed in appendix \ref{app:FIR}.

\section{Results}\label{sec.results}
\begin{figure}[!htb]
	\includegraphics[scale=0.48]{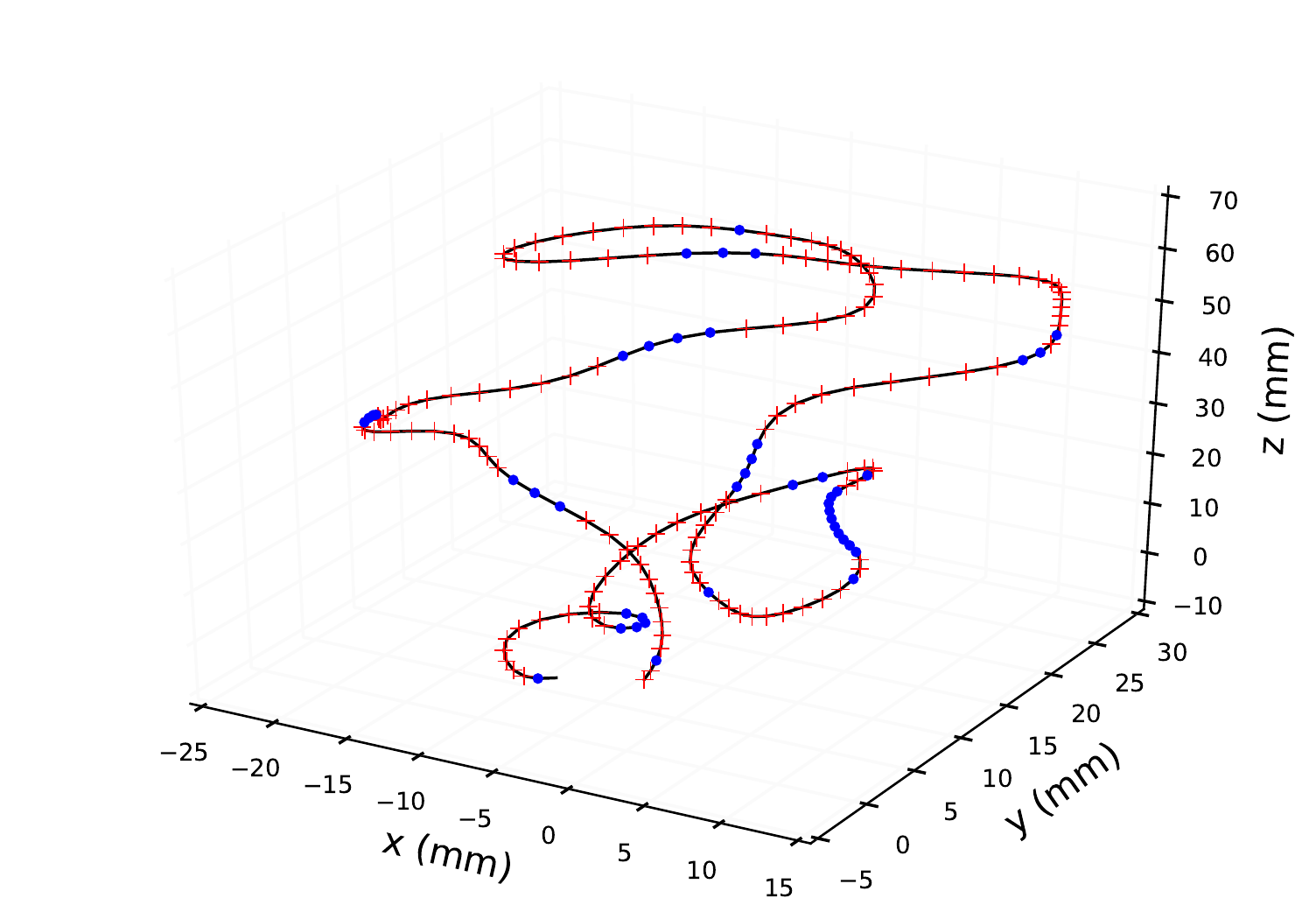}
	\includegraphics[scale=0.48]{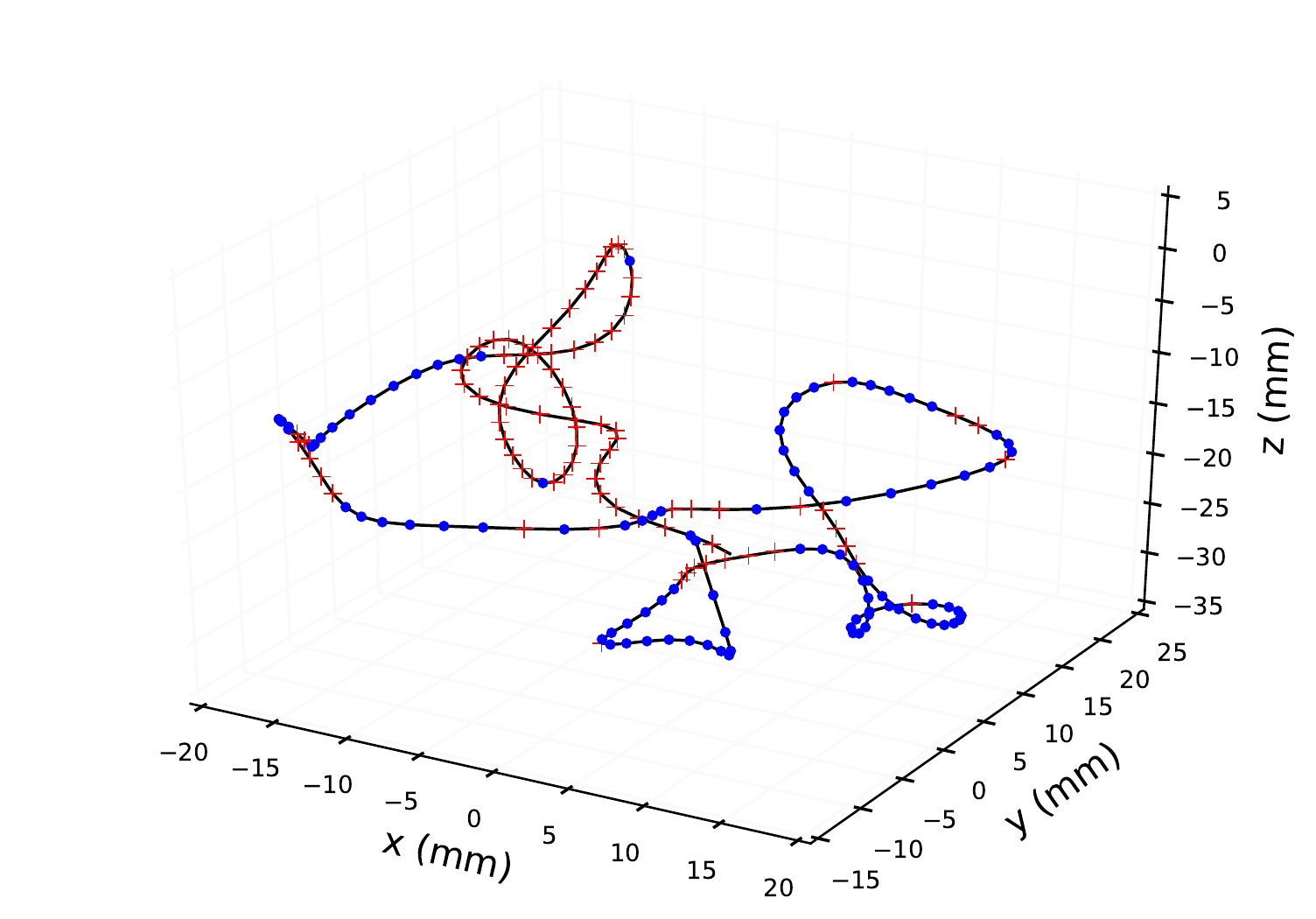}
	\caption{\label{fig_trkcurv}Calculated curvature sign at each point along the track for a single-electron event (left) and a double-beta event (right) generated using the GEANT-4 based Monte Carlo.  The red $+$ markers indicate positive curvature, while the blue dots indicate negative curvature.  Note that a lowpass filter has been applied to the tracks shown as described in appendix \protect\ref{app:FIR}.}
\end{figure}

The analysis described in section \ref{sec.curvature} was performed for single-electron and
\bbonu\ tracks for all combinations of $P =$ 5, 10, and 15 atm and $B =$ 0.1, 0.3, 0.5, 0.7, 
and 1.0 T.  In all of these configurations, the hits were smeared in $(x,y,z)$ with $\sigma_{s} = 2$ mm.  They were also ``sparsed,'' that is, groups of every $N_{s}$ hits were averaged into a single hit, with $N_{s} = 2$.  Additional analyses in the $P =$ 10 atm, $B = 0.5$ T configuration were performed, one with $\sigma_{s} = 1$ mm and $N_{s} = 1$, and another with $\sigma_{s} = 3$ mm and $N_{s} = 3$.  For each configuration, $10^4$ events of each track type (single-electron and \bbonu) were generated. 
For illustrative purposes, figure \ref{fig_trkcurv} shows the calculated curvature signs for a single-electron and double-beta event, after filtering with a lowpass filter, at $B = 0.5$ T and $P = 10$ atm.  

\begin{figure}[!htb]
	\includegraphics[scale=0.44]{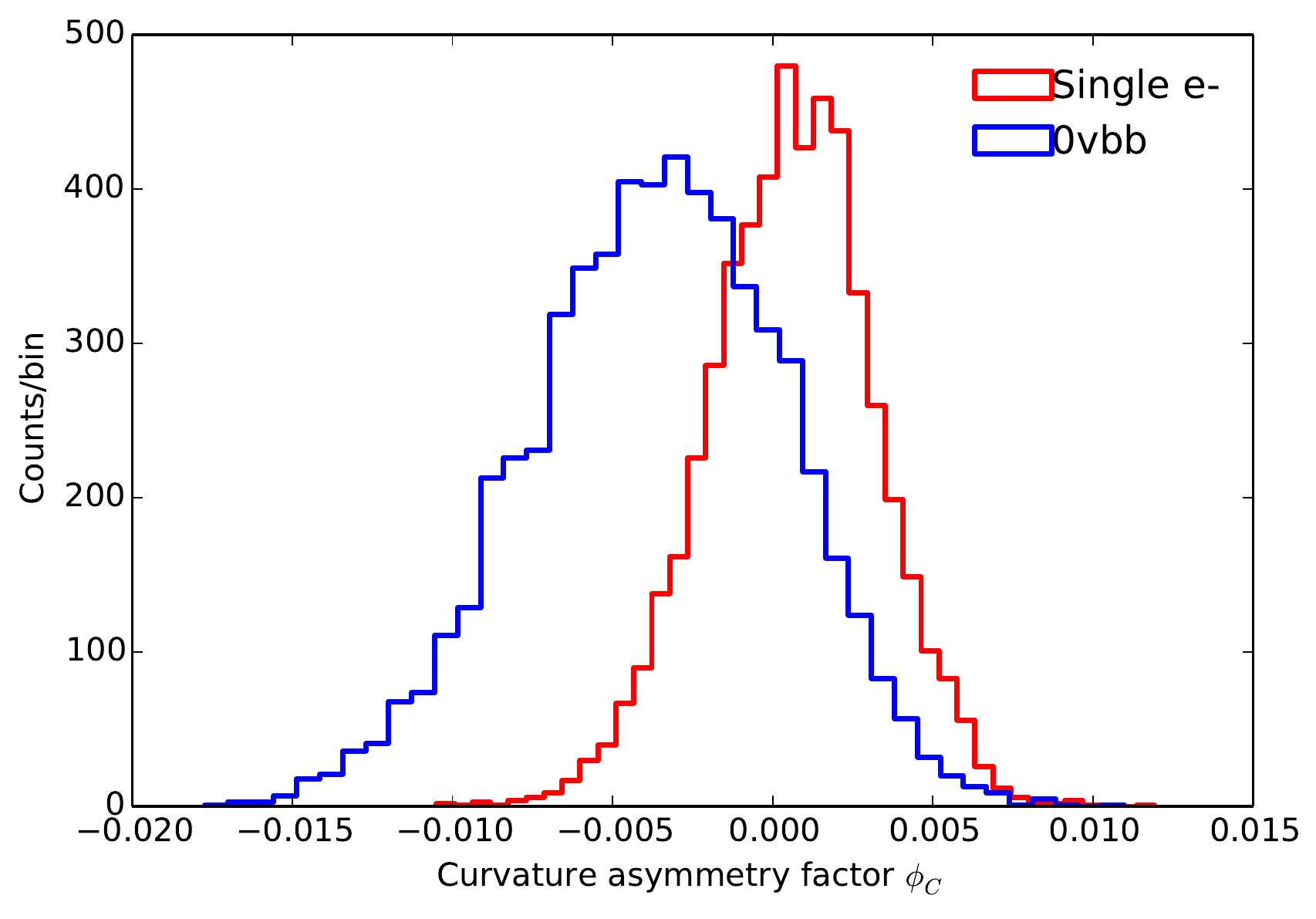}
	\includegraphics[scale=0.44]{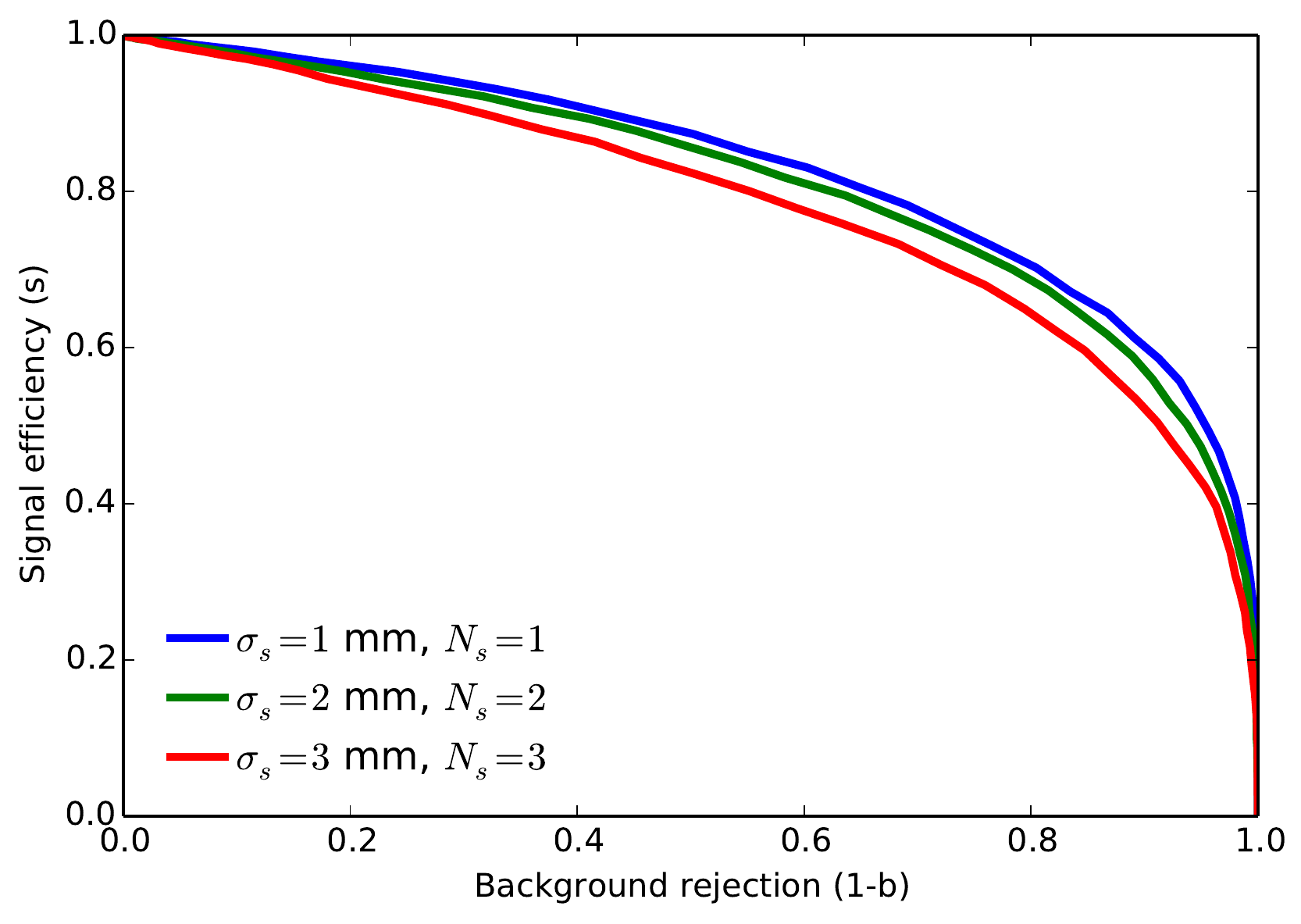}
	\caption{\label{fig_svsbg}Curvature asymmetry factor for single-electron and \bbonu\ events with $\sigma_{s} = 2$ mm  (left) and resulting signal efficiency vs. background rejection curve produced by varying a cut on $\phi_{C}$ space, shown also for $\sigma_{s} = 1$ mm and $\sigma_{s} = 3$ mm (right).  Here $s$ is the fraction of the total number of \bbonu\ events considered that was identified as a \bbonu\ event, and $b$ is the fraction of the total number of single-electron background events considered that was identified as a \bbonu\ event.  These results were obtained for the $P = 10$ atm and $B = 0.5$ T configuration.}
\end{figure}

For each track we calculate the curvature asymmetry factor shown in equation \ref{eqn_assym}. Next, a cut was placed on the CAF, defining the events considered to be candidate 
\bbonu\ tracks.  The cut was varied, and in each case 
the fraction of single-electron (background) tracks rejected by the cut was determined along with the fraction of 
\bbonu\ (signal) tracks accepted.\footnote{Of the $10^4$ events generated for each configuration and track type, only events that passed a set of initial topological cuts were considered in the curvature analysis.  These cuts required a minimum total deposited energy of 2.4 MeV and the deposition of this energy in a single continuous track, so events with some energy deposited significantly away ($> 2$ cm) from the main track were not analyzed further.  The final fraction of events therefore refers to a fraction of this subset of events analyzed, not a fraction of the total $10^4$ events generated.}  An example of such an analysis is summarized in figure \ref{fig_svsbg}.

\begin{figure}[!htb]
	\centering
	\includegraphics[scale=0.6]{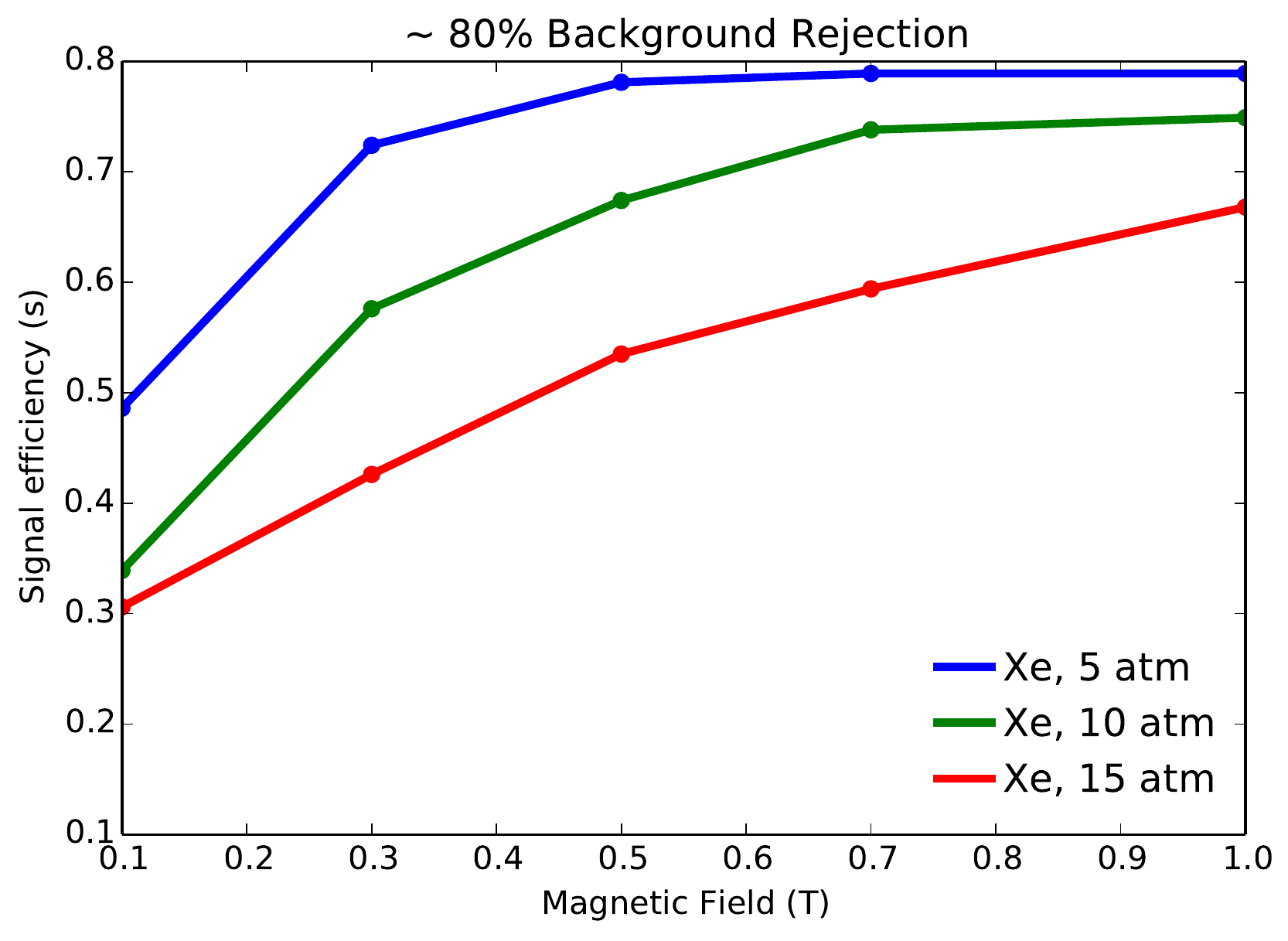} \\
	\includegraphics[scale=0.6]{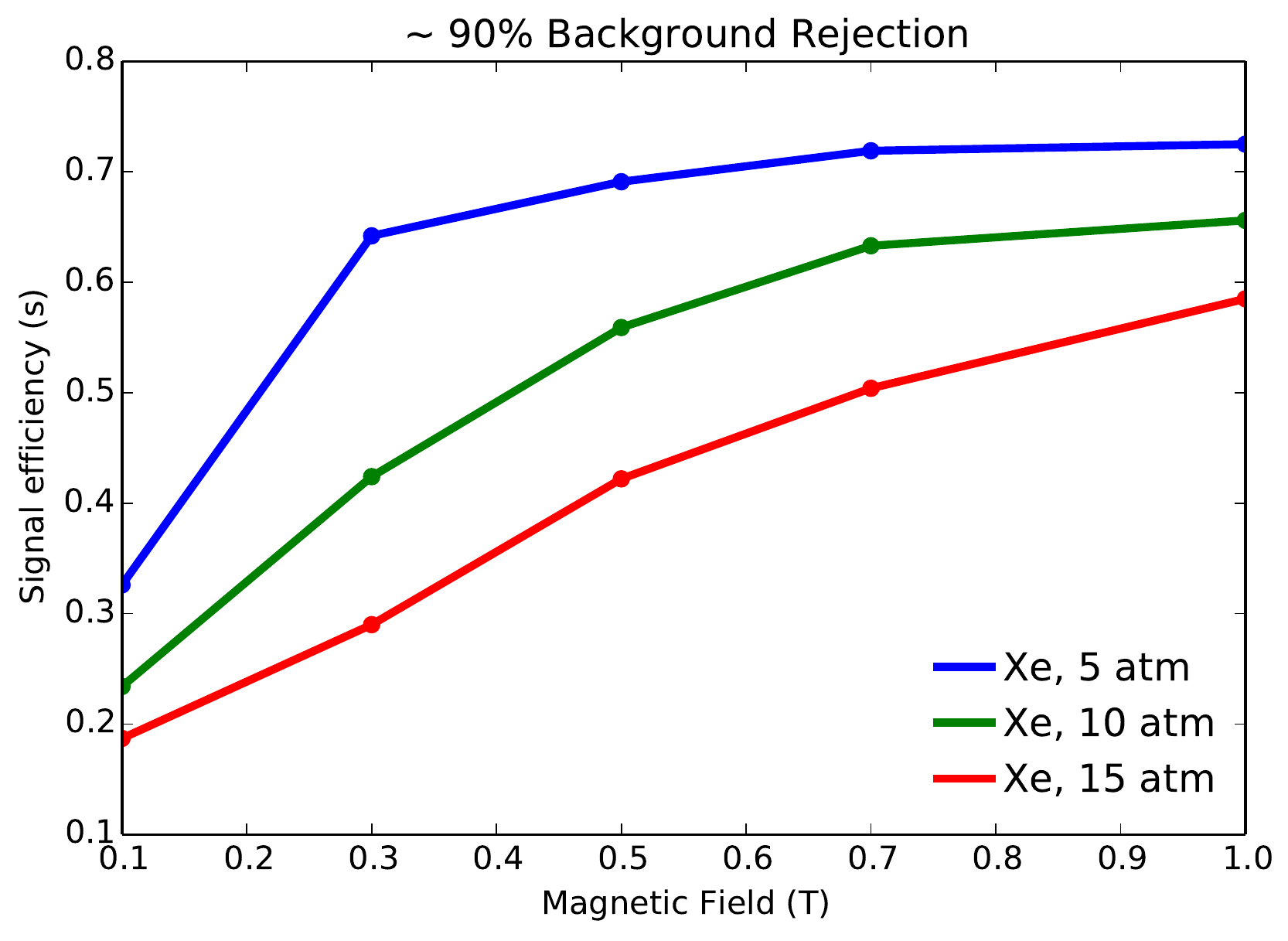}
	\caption{\label{fig_config}Signal efficiency corresponding to approximately 80\% background rejection (above) and 90\% background rejection (below) vs. magnitude of applied external magnetic field for different pressures.}
\end{figure}

To illustrate the performance of the method at the different configurations of gas pressure and magnetic 
field, we examine the signal efficiency $s$ (fraction of \bbonu\ events correctly identified) obtained 
given a cut that provides approximately 80\% and 90\% background rejection.  The results are shown in figure 
\ref{fig_config}.

Note that the results depend significantly of the operating pressure. The best performance is obtained, as expected, at the lowest considered pressure (5 bar), and it reaches a plateau (80\% efficiency for a background rejection of 80\%, and 70\% efficiency for a background rejection of 90\%) for magnetic fields larger than 0.5 Tesla. The performance at 10 bar appears also quite acceptable. It reaches a plateau (75\% efficiency for a background rejection of 80\%, and 65\% efficiency for a background rejection of 90\%) for magnetic fields larger than 0.7 Tesla. The performance at 15 bar is poorer and does not saturate with the magnetic field, suggesting that some additional improvement can be obtained with larger values of $B$. 

\section{An improved analysis}\label{sec.improvedanalysis}
\begin{figure}[!htb]
	\centering
	\includegraphics[scale=0.55]{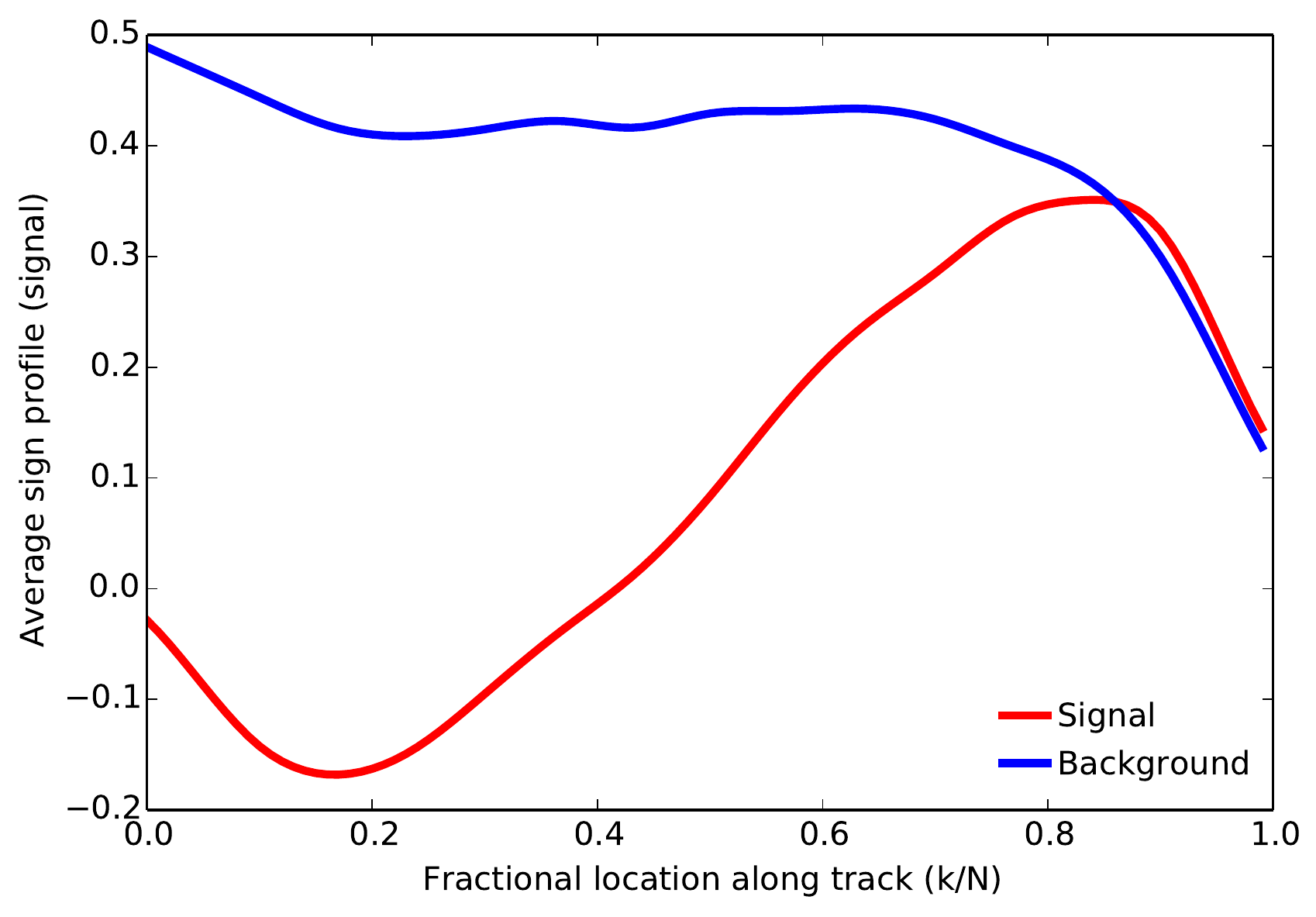}
	\caption{\label{fig_profiles}Average curvature sign profiles for single-electron and $0\nu\beta\beta$ events generated at $B = 0.5$ T and $P = 10$ atm, with $\sigma_{s} = 2$ mm and $N_{s} = 2$.  The profile shows the average sign vs. fractional distance along the track ($k/N$, where $k$ is the hit number and $N$ is the total number of hits).  The profiles were created by dividing the normalized track into 15 bins, and for each calculated curvature value for each event, adding a value of $+1$ or $-1$ to the bin, then dividing each bin by the total number of values placed in it.  The values in the first and second bin were extrapolated linearly to $k/N = 0$, and those in the final two bins were extrapolated to $k/N = 1$.  The entire set of values was then interpolated using cubic polynomials.}
\end{figure}

While the curvature asymmetry factor $\phi_C$ from equation \ref{eqn_assym} provides a reasonable separation between signal and background events, it still represents a rather minimal approach.  For example, in calculating $\phi_C$, the difference was taken in the average of the signs over the two halves of the track, but the vertex in a $0\nu\beta\beta$ event is not likely to be found in the center.  A more detailed analysis taking into account more features of the two different classes of events should give improved performance.  As an example of such an analysis, we consider constructing a track ``profile'' for each type of event containing the value of the curvature sign at each point along the track averaged over all events.  Since not all tracks will be of exactly the same length, we normalize the track length to 1, and thus the profile contains an average sign value for each fractional distance along the track.  Note that fractional distance is determined by hit number $k$ divided by the total number of hits $N$ rather than actual distance traveled.  Figure \ref{fig_profiles} shows profiles made for simulated single-electron and $0\nu\beta\beta$ events.  The profiles show a clear separation between signal and background events earlier in the track where the sign of the curvature is most likely to differ.

Once the profiles have been generated, they can be used to define a new variable that will provide separation between the two classes of events.  For each event, the sign of the curvature $\mathrm{sgn}(\kappa_{i})$ at each (normalized) distance along the track is compared to that of the profile for signal ($0\nu\beta\beta$) events, $p_{S}$, and background (single-electron) events, $p_{B}$, in creating two $\chi^2$-like sums,

\begin{equation}
 \chi^2_{S} = \sum_{k=0}^{N}\frac{[\mathrm{sgn}(\kappa_{k}) - p_{S}(k/N)]^{2}}{\sigma^2_{s}(k/N)},
\end{equation}

\noindent and a similar sum $\chi^2_B$ in which the background profile $p_{B}$ is used in place of $p_{S}$.  The values $p_S(x)$ and $p_B(x)$ are the mean curvature signs at fractional track location $x = k/N$, and the values $\sigma^2_S$ and $\sigma^2_B$ are the variances of the mean curvature sign.  The variances were found to be near 1 at all $k/N$ in both profiles, reflecting the large amount of fluctuation in curvature sign at any given point on the track due to multiple scattering.  Despite these large variances, the large number of events used in determining the average values made for low errors on the mean and thus smooth profiles.  The discriminating variable is then $r_{\chi} = \chi^2_B/\chi^2_{S}$.  For background events, this ratio should be smaller, as the curvature signs throughout the track should be more similar to the background event profile (smaller $\chi^2_B$) and more different than the signal event profile (larger $\chi^2_S$).  Based on similar arguments, this ratio should be larger for signal events.

\begin{figure}[!htb]
	\centering
	\includegraphics[scale=0.55]{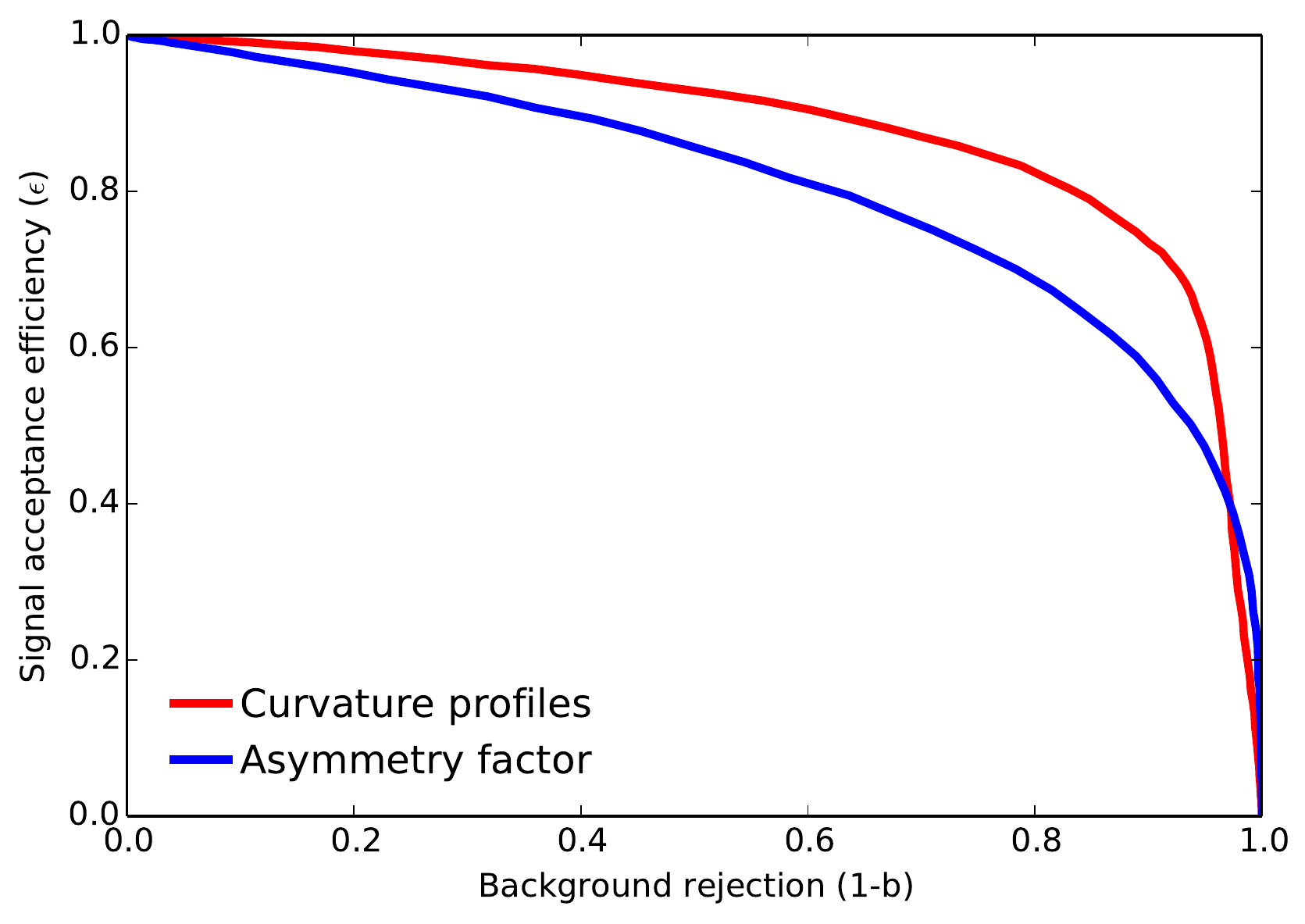}
	\caption{\label{fig_svsbprof}Signal acceptance efficiency vs. background rejection obtained at $B = 0.5$ T and $P = 10$ atm, with $\sigma_{s} = 2$ mm and $N_{s} = 2$.  The curve obtained using the curvature asymmetry factor is shown along with that obtained using the profile-based method.}
\end{figure}

Figure \ref{fig_svsbprof} compares the signal efficiency obtained for a given background rejection using the profile-based method (discriminant variable $r_{\chi}$) described in this section and the method based on the asymmetry factor $\phi_{C}$ described in section \ref{sec.curvature}.  The profile-based method performs better up to a background rejection of about 95\%.  This clearly shows the potential for further optimization. In our ``standard test case'' (10 atm, 2 mm resolution, 0.5 Tesla) a rejection factor of one order of magnitude is achieved for a signal efficiency of about 75\% using the profile method. 

\section{Outlook: the case of a ton-scale HPXe TPC with a magnetic field}\label{sec.outlook}

The application of an external magnetic field in a high-pressure detector capable of accurate particle track reconstruction with resolution of approximately 2 mm in $(x,y,z)$ can provide an additional background rejection factor of 90\% with good signal efficiency (70-80\%) at pressures of about 10 atm.  The background rejection improves with decreasing pressure due to the more extended tracks produced at lower gas densities.  A sufficient track reconstruction algorithm has not yet been realized to take full advantage of these findings, and the NEXT collaboration is currently working towards a potentially suitable algorithm.

It follows that the background rejection of a future ton-scale HPXe using a magnetic field could be up to a factor of 10 better than the current rejection factor expected by the NEXT-100 experiment (while keeping an efficiency in the range of 25\%). The resulting background rate would be around $5 \cdot 10^{-5}$ \ckky\ or about a half count per ton and year in the ROI, assuming a resolution of 0.5\% FWHM at \Qbb. This is enough, for most NME sets, to fully cover the inverse hierarchy  in a few years. 

\appendix

\section{A Lowpass FIR Filter}\label{app:FIR}

One way to make the calculation of the curvature of a track less susceptible to noise 
introduced through multiple scattering and non-ideal reconstruction resolution is to apply a lowpass filter to 
the lists of values for each coordinate x, y, and z.  We now look at the arrays of x, y, and z coordinates as digital 
signals in the time domain, e.g. $x[n]$, where $n$ is the index of the corresponding hit.  Each array can be 
represented in the frequency domain $X[k]$ using the discrete Fourier transform

\begin{equation}
X[k] = \sum_{n=0}^{N-1}x[n]e^{-i2\pi kn/N},
\end{equation}

\noindent where $N$ is the total number of samples and $k$ is the discrete frequency of each complex
sinusoid $e^{-i2\pi kn/N}$ in cycles per $N$ samples.  This discrete frequency can be translated to an analog
frequency $f_{k}$ (for example in units of time$^{-1}$) by knowing the frequency at which the digital signal was
sampled, or the sampling frequency $f_{s}$ in samples per unit time, as

\begin{equation}
f_{k} = kf_{s}.
\end{equation}

The goal is to apply a digital lowpass filter to the coordinate arrays that serves to
smooth the track by eliminating high-frequency noise yet retains the curvature of the track
due to the magnetic field.  The ideal lowpass filter will serve to eliminate sinusoidal components
$X[k]$ in the digital signal for $k$ greater than some cutoff frequency $k_{c}$ and allow 
others with $k$ less than $k_{c}$.  In practice, the filter will have a non-ideal stopband in which sinusoidal 
components with frequencies less than $k_{c}$ will be increasingly preserved with decreasing $k$ and those 
with frequencies greater than $k_{c}$ will be increasingly attenuated with increasing $k$.  The filter must be 
designed to ensure we don't eliminate the sinusoidal motion introduced by the magnetic field, and we use a
filter with a wide stopband (this reduces its complexity) and place $k_{c}$ near the discrete frequency 
corresponding to the cyclotron frequency

\begin{equation}\label{eqn_wcyc}
\omega_{\mathrm{cyc}} = qB/m_{e} = 1.76B \times 10^{11} \,\, \mathrm{rad/s},
\end{equation}

\noindent where $q \approx -1.60 \times 10^{-19}$ is the electron charge in Coulombs, $B$ is the magnetic field
strength in Tesla, and $m_{e} \approx 9.11 \times 10^{-31}$ is the electron mass in kg.  The filter will be a finite impulse response (FIR) filter and will take the form of a  list of coefficients $b_{m}$.  It is applied to the signal $x[n]$ as

\begin{equation}
x_{f}[n] = \sum_{m=0}^{N_t} b_{m}x[n-m]
\end{equation}

\noindent where the number of filter coefficients $N_{t}$ is the order of the filter.  A delay will exist in the
final filtered signal equal to $N_{t}/2$ samples, and the final $N_{t}/2$ samples will not be useful in the final analysis and are removed.
For a track sampled in x, y, and z for a total of $N$ samples, to determine what discrete frequency $k_{cyc}$ to which this analog frequency ($\omega_{\mathrm{cyc}}/2\pi$ in cycles per second) corresponds, one must know the frequency with which the track has been sampled.  An average track production time $\overline{T}$ can be calculated from tabulated values of $dE/dx$ (which, since this refers to kinetic energy loss, we will call $dK/dx$) as

\begin{equation}\label{eqn_T}
\overline{T} = \frac{1}{c}\int_{0}^{K_{f}} \biggl[\frac{\sqrt{(K^2-m_e^2)}}{K+m_e}(dK/dx)\biggr]^{-1} dK \approx 1.25 \,\, \mathrm{ns},
\end{equation}

\noindent where here we have used $K_{f} = Q_{\beta\beta} = 2.447$ MeV and tabulated 
$dE/dx$ from NIST \cite{NIST_mac} in xenon.  The sampling frequency is then $f_{s} = N/T$, and so the motion
due to the magnetic field should manifest itself in the arrays of sampled x and y coordinates
of the track as a sinusoidal component of discrete frequency 

\begin{equation}\label{eqn_kcyc}
k_{\mathrm{cyc}} = (qB/m)\cdot(\overline{T}/N).
\end{equation}

\begin{figure}[!htb]
	\centering
	\includegraphics[scale=0.6]{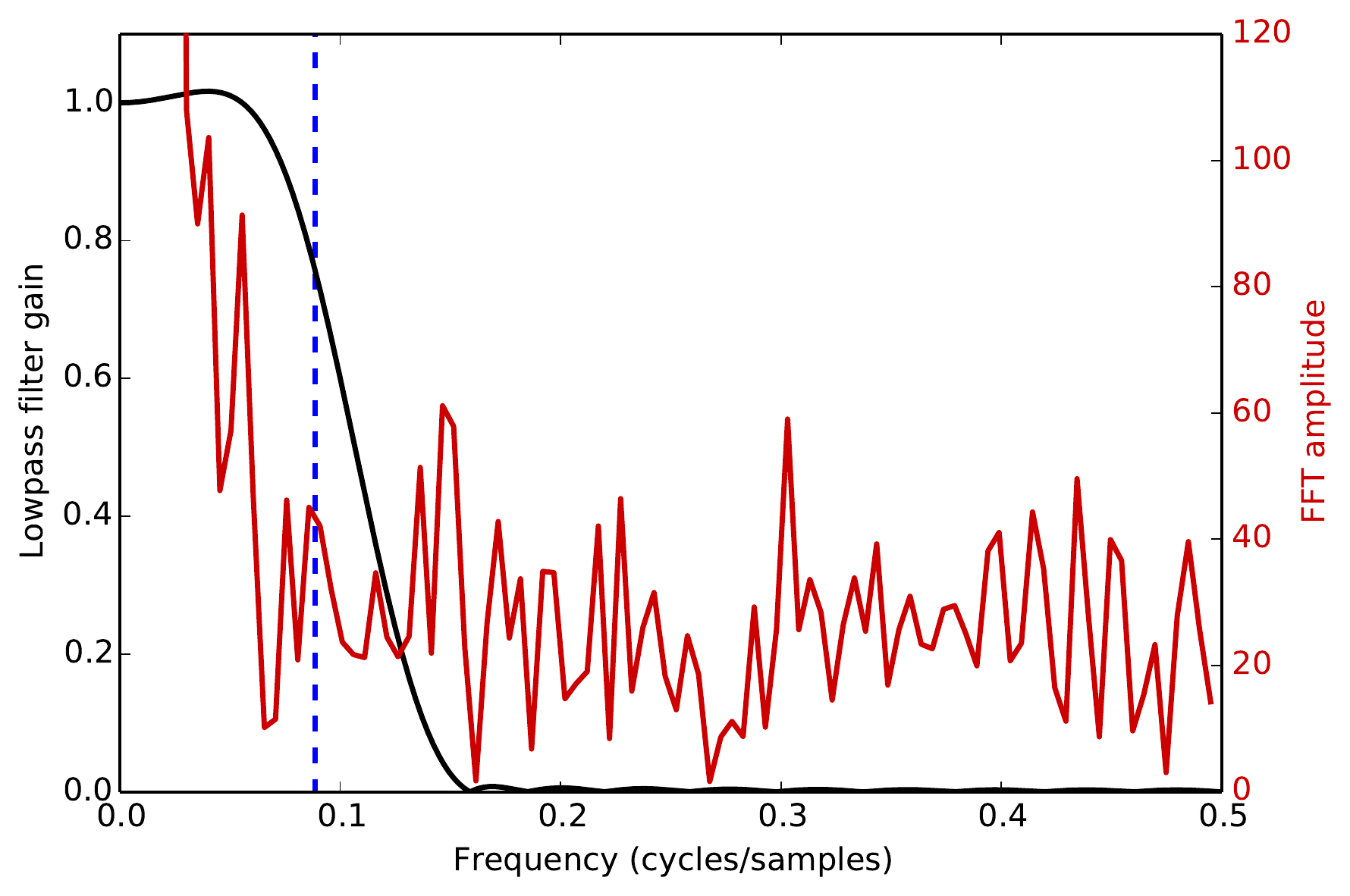}
	\caption{\label{fig_FIR}FIR filter gain vs. discrete frequency $k$ (solid black curve) designed for a track of length $N = 198$ samples.  The coefficients $X[k]$ of the fast fourier transform (FFT) of the x-coordinate array of the track are shown vs.\ discrete frequency in red.  The cutoff frequency used in designing the filter is shown as a blue vertical line.  The filter serves to eliminate the high-frequency noise components of the track that make the calculation of derivatives more difficult.}
\end{figure}

\noindent One should ensure that the track is sampled at a rate higher than two times the cyclotron
frequency, that is $N/T > \omega_{\mathrm{cyc}}/2\pi$, or the helical motion will not be properly
reconstructed.  An example filter is shown in figure \ref{fig_FIR}.  

For the track shown in figure \ref{fig_flt}, an FIR filter was designed with 
cutoff frequency equal to $k_c = 1.2k_{\mathrm{cyc}}$ and stopband width equal to 0.2.  This was sufficient to 
eliminate high-frequency noise, leaving a smooth track for the calculation of derivatives, yet the filter order was
low enough so as to not lose a significant number of the track samples.  For this track, a number of samples 
$N = 198$ and using $T = 1.25$ ns from equation \ref{eqn_T} and $\omega_{\mathrm{cyc}} = 0.88 \times 
10^{11}$ rad/s from equation \ref{eqn_wcyc} with $B = 0.5$ T, we have $k_{\mathrm{cyc}}/2\pi = 0.089$ 
cycles/sample as shown in figure \ref{fig_FIR}.

\begin{figure}[!htb]
	\centering
	\includegraphics[scale=0.6]{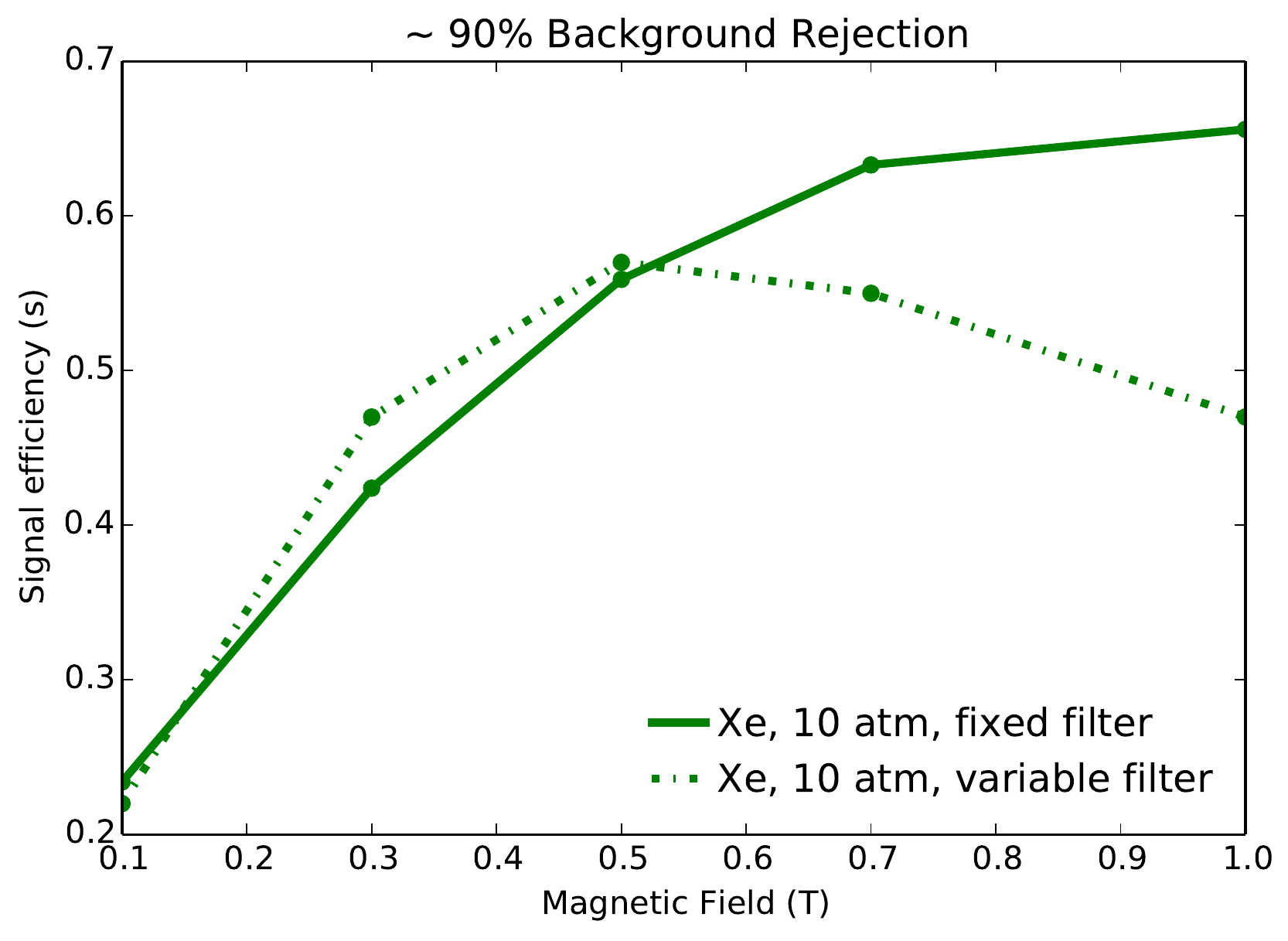}
	\caption{\label{fig_FIRdependence}Signal efficiency vs. magnetic field for 90\% background rejection shown with two different filter calculations to demonstrate the effect of the lowpass filter design on the background rejection performance.  The use of a filter with cutoff frequency calculated on a track-by-track basis (``variable filter'') does not yield a remarkable improvement in performance at low values of $B$ and allows too much noise to remain at high values of $B$.  The use of a filter with a fixed cutoff frequency (``fixed filter,'' same as Xe 10 atm curve in figure \protect\ref{fig_config} left) in the range of those calculated for $B = 0.5$ T for all tracks eliminates the noise but does not eliminate information about the curvature to the extent that the performance is degraded.}
\end{figure}

As the magnitude of $B$ increases, the frequency of helical motion often becomes great enough that placing the 
filter cutoff frequency $k_{c}$ near the cyclotron frequency does not permit sufficient filtering of the 
high-frequency components of the track, and the resulting track is not smooth enough for a clean calculation of 
the curvature.  Therefore it may be an advantage to place the cutoff frequency of the filter below 
$k_{\mathrm{cyc}}$ despite increased attenuation at the discrete frequency corresponding to the helical motion 
induced by the magnetic field.\footnote{In any case, due to the constant diversion from the ideal path of the electron
introduced by multiple scattering, the curvature of the track in the magnetic field will not appear at a single 
distinct value in the frequency domain.}  Figure \ref{fig_FIRdependence} shows the signal efficiency vs. magnetic
field for a background rejection of 90\% at 10 atm in pure xenon for which the filter was designed for each track 
individually with cutoff frequency equal to $1.2k_{\mathrm{cyc}}$, and for which the cutoff frequency of all filters 
was chosen to be a fixed value.

Because the fixed-cutoff analysis performs significantly better at higher fields, it was chosen as the primary 
analysis in this study.  For the results shown in section \ref{sec.results}, the lowpass filter applied to all tracks 
was designed with a fixed cutoff frequency of $k_{c}/2\pi = 0.102$ (with the exception of those shown in figure \ref{fig_trkcurv}, to which the filter shown in figure \ref{fig_FIR} specifically was applied).

\acknowledgments

This work was supported by the European Research Council under the Advanced Grant 339787-NEXT and the Ministerio de Econom\'{i}a y Competitividad of Spain under Grants CONSOLIDER-Ingenio 2010 CSD2008-0037 (CUP), FPA2009-13697-C04-04, FPA2009-13697-C04-01, FIS2012-37947-C04-01, FIS2012-37947-C04-02, FIS2012-37947-C04-03, and FIS2012-37947-C04-04.  JJGC would like to acknowledge the Aspen Physics Center and the Simons Foundation for their support.

\bibliography{nextb}

\end{document}